# Heat Waves – A hot topic in climate change research

Werner Marx*+, Robin Haunschild*, Lutz Bornmann*,**

* Max Planck Institute for Solid State Research
Heisenbergstr. 1, 70569 Stuttgart, Germany
Email: w.marx@fkf.mpg.de; r.haunschild@fkf.mpg.de

** Administrative Headquarters of the Max Planck Society
Science Policy and Strategy Department
Hofgartenstr. 8, 80539 Munich, Germany
Email: bornmann@gv.mpg.de

+ Corresponding author

ORCID-ID Werner Marx: https://orcid.org/0000-0002-1763-5753

ORCID-ID Robin Haunschild: https://orcid.org/0000-0001-7025-7256

ORCID-ID Lutz Bornmann: https://orcid.org/0000-0003-0810-7091

## Abstract

Research on heat waves (periods of excessively hot weather, which may be accompanied by high humidity) is a newly emerging research topic within the field of climate change research with high relevance for the whole of society. In this study, we analyzed the rapidly growing scientific literature dealing with heat waves. No summarizing overview has been published on this literature hitherto. We developed a suitable search query to retrieve the relevant literature covered by the Web of Science (WoS) as complete as possible and to exclude irrelevant literature (n = 8,011 papers). The time-evolution of the publications shows that research dealing with heat waves is a highly dynamic research topic, doubling within about 5 years. An analysis of the thematic content reveals the most severe heat wave events within the recent decades (1995 and 2003), the cities and countries/regions affected (United States, Europe, and Australia), and the ecological and medical impacts (drought, urban heat islands, excess hospital admissions, and mortality). Risk estimation and future strategies for adaptation to hot weather are major political issues. We identified 104 citation classics which include fundamental early works of research on heat waves and more recent works (which are characterized by a relatively strong connection to climate change).

## 1 Introduction

As a consequence of the well-documented phenomenon of global warming, climate change has become a major research field in the natural and medical sciences, and more recently also in the social and political sciences. The scientific community has contributed extensively to a comprehensive understanding of the earth's climate system, providing various data and projections on the future climate as well as on the effects and risks of anticipated global warming (IPCC 2014; CSSR 2017; NCA4 2018; and the multitude of references cited therein). During recent decades, climate change has also become a major political, economic, and environmental issue and a central theme in political and public debates.

One consequence of global warming is the increase of extreme weather events such as heat waves, droughts, floods, cyclones, and wildfires. Some severe heat waves occurring within the last few decades made heat waves a hot topic in climate change research, with "hot" having a dual meaning: high temperature and high scientific activity. "More intense, more frequent, and longer lasting heat waves in the 21$^{st}$ century" is the title of a highly cited paper published 2004 in *Science* (Meehl and Tebaldi 2004). This title summarizes in short what most climate researchers anticipate for the future. But what are heat waves (formerly also referred to as "heatwaves")? In general, a heat wave is a period of excessively hot weather, which may be accompanied by high humidity. Since heat waves vary according to region, there is no universal definition, but only definitions relative to the usual weather in the area and relative to normal temperatures for the season. The World Meteorological Organization (WMO) defines a heat wave as five or more consecutive days of prolonged heat in which the daily maximum temperature is higher than the average maximum temperature by 5 °C (9 °F) or more (https://www.britannica.com/science/heat-wave-meteorology).

Europe, for example, has suffered from a series of intense heat waves since the beginning of the 21$^{st}$ century. According to the World Health Organization (WHO) and various national reports, the extreme 2003 heat wave caused about 70,000 excess deaths, primarily in France and Italy. The 2010 heat wave in Russia caused extensive crop loss, numerous wildfires, and about 55,000 excess deaths (many in the city of Moscow). Heat waves typically occur when high pressure systems become stationary and the winds on their rear side continuously pump hot and humid air northeastward, resulting in extreme weather conditions. The more intense and more frequently occurring heat waves cannot be explained solely by natural climate variations and without human made climate change (IPCC 2014; CSSR 2017; NCA4 2018). Scientists discuss a weakening of the polar jet stream caused by global warming as a possible reason for an increasing probability for the occurrence of stationary weather, resulting in heavy rain falls or heat waves (Broennimann et al. 2009; Coumou et al. 2015; Mann 2019). This jet stream is one of the most important factors for the weather in the middle latitude regions of North America, Europe, and Asia.

Until the end of the 20th century, heat waves were predominantly seen as a recurrent meteorological fact with major attention to drought, being almost independent from human activities and unpredictable like earthquakes. However, since about 1950, distinct changes in extreme climate and weather events have been increasingly observed. Meanwhile, climate change research has revealed that these changes are clearly linked to the human influence on the content of greenhouse gases in the earth's atmosphere. Climate-related extremes, such as heat waves, droughts, floods, cyclones, and wildfires, reveal significant vulnerability to climate change as a result of global warming.

In recent years, research on heat waves has been established as an emerging research topic within the large field of current climate change research. Bibliometric analyses are very suitable in order to have a systematic and quantitative overview of the literature that can be assigned to an emerging topic such as research dealing with heat waves (e.g., Haunschild et al. 2016). No summarizing overview on the entire body of heat wave literature has been published until now. However, a bibliometric analysis of research on urban heat islands as a more specific topic in connection with heat waves has been performed (Huang and Lu 2018).

In this study, we analyzed the publications dealing with heat waves using appropriate bibliometric methods and tools. First, we determined the amount and time-evolution of the scientific literature dealing with heat waves. The countries contributing the most papers are presented. Second, we analyzed the thematic content of the publications via keywords assigned by the WoS. Third, we identified the most important (influential) publications (and also the historical roots). We identified 104 citation classics, which include fundamental early works and more recent works with a stronger connection to climate change.

## 2 Heat waves as a research topic

The status of the current knowledge on climate change is summarized in the *Synthesis Report of the Fifth Assessment Report (AR5)* of the *Intergovernmental Panel on Climate Change (IPCC)* (IPCC 2014, https://www.ipcc.ch/report/ar5/syr/). This panel is the United Nations body for assessing the science related to climate change. The *Synthesis Report* is based on the reports of the three *IPCC Working Groups*, including relevant *Special Reports*. In its *Summary for Policymakers* it provides an integrated view of climate change as the final part of the *Fifth Assessment Report* (IPCC 2014, https://www.ipcc.ch/site/assets/uploads/2018/02/AR5_SYR_FINAL_SPM.pdf).

In the chapter *Extreme Events*, it is stated that "changes in many extreme weather and climate events have been observed since about 1950. Some of these changes have been linked to human influences, including a decrease in cold temperature extremes, an increase in warm temperature extremes, an increase in extreme high sea levels and an increase in the number of heavy precipitation events in a number of regions … It is very likely that the number of cold days and nights has decreased and the number of warm days and nights has increased on the global scale. It is likely that the frequency of heat waves has increased in large parts of Europe, Asia and Australia. It is very likely that human influence has contributed to the observed global scale changes in the frequency and intensity of daily temperature extremes since the mid-20$^{th}$ century. It is likely that human influence has more than doubled the probability of occurrence of heat waves in some locations" (p. 7-8). Under *Projected Changes*, the document summarizes as follows: "Surface temperature is projected to rise over the 21$^{st}$ century under all assessed emission scenarios. It is very likely that heat waves will occur more often and last longer, and that extreme precipitation events will become more intense and frequent in many regions" (p. 10).

With regard to the United States, the *Climate Science Special Report* of the *U.S. Global Change Research Program* (CSSR 2017, https://science2017.globalchange.gov/) mentions quite similar observations and states unambiguously in its *Fourth National Climate Assessment (Volume I)* report (https://science2017.globalchange.gov/downloads/CSSR2017_FullReport.pdf) under *Observed Changes in Extremes* that "the frequency of cold waves has decreased since the early 1900s, and the frequency of heat waves has increased since the mid-1960s (very high confidence). The frequency and intensity of extreme heat and heavy precipitation events are increasing in most continental regions of the world (very high confidence). These trends are consistent with expected physical responses to a warming climate [p. 19]. Heavy precipitation events in most parts of the United States have increased in both intensity and frequency since 1901 (high confidence) [p. 20]. There are important regional differences in trends, with the largest increases occurring in the northeastern United States (high confidence). Recent droughts and associated heat waves have reached record intensity in some regions of the United States … (very high confidence) [p. 21]. Confidence in attribution findings of anthropogenic influence is greatest for extreme events that are related to an aspect of temperature" (p. 123).

Among the key findings in the chapter on temperature changes in the United States the report states that "there have been marked changes in temperature extremes across the contiguous United States. The frequency of cold waves has decreased since the early 1900s, and the frequency of heat waves has increased since the mid-1960s (very high confidence). Extreme temperatures in the contiguous United States are projected to increase even more than average temperatures. The temperatures of extremely cold days and extremely warm days are both expected to increase. Cold waves are projected to become less intense while heat waves will become more intense (very high confidence) [p. 185]. Most of this methodology as applied to extreme weather and climate event attribution, has evolved since the European heat wave study of Stott et al." (p. 128).

Heat waves are also discussed in the *Fourth National Climate Assessment (Volume II)* report (NCA4 2018, https://nca2018.globalchange.gov/). The *Report-in-Brief* (https://nca2018.globalchange.gov/downloads/NCA4_Report-in-Brief.pdf) for example states: "More frequent and severe heat waves and other extreme events in many parts of the United States are expected [p. 38]. Heat waves and heavy rainfalls are expected to increase in frequency and intensity [p. 93]. The season length of heat waves in many U.S. cities has increased by over 40 days since the 1960s [p. 30]. Cities across the Southeast are experiencing more and longer summer heat waves [p. 123]. Exposure to hotter temperatures and heat waves already leads to heat-associated deaths in Arizona and California. Mortality risk during a heat wave is amplified on days with high levels of ground-level ozone or particulate air pollution" (p. 150).

In summary, climate change research expects more frequent and more severe heat wave events as a consequence of global warming. It is likely that the more frequent and longer lasting heat waves will significantly increase excess mortality, particularly in urban regions with high air pollution. Therefore, research around heat waves will become increasingly important and is much more than a temporary research fashion.

## 3 Methodology

### 3.1 Data set used

This analysis is based on the relevant literature retrieved from the following databases accessible under the Web of Science (WoS) of Clarivate Analytics: Web of Science Core Collection: Citation Indexes, Science Citation Index Expanded (SCI-EXPANDED), Social Sciences Citation Index (SSCI), Arts & Humanities Citation Index (A&HCI), Conference Proceedings Citation Index- Science (CPCI-S), Conference Proceedings Citation Index- Social Science & Humanities (CPCI-SSH), Book Citation Index– Science (BKCI-S), Book Citation Index– Social Sciences & Humanities (BKCI-SSH), Emerging Sources Citation Index (ESCI).

We applied the search query given in the appendix to cover the relevant literature as completely as possible and to exclude irrelevant literature. We practiced an iterative query optimization by identifying and excluding the WoS subject categories with most of the non-relevant papers. For example, heat waves are also mentioned in the field of materials science but have nothing to do with climate and weather phenomena. Unfortunately, WoS obviously assigned some heat wave papers related to climate to materials science related subject categories. Therefore, these subject categories were not excluded. By excluding the other non-relevant subject categories, 597 out of 8,568 papers have been removed, resulting in a preliminary publication set of 7,971 papers (#2 of the search query). But this is no safe method, since the excluded categories may well include some relevant papers. Therefore, we have combined these 597 papers with search terms related to climate or weather and retrieved 62 relevant papers in addition, which we added to our preliminary paper subset, eventually receiving 8,033 publications (#3 to #5 of the search query).

Commonly, publication sets for bibliometric analyses are limited to articles, reviews, and conference proceedings as the most relevant document types and are restricted to complete publication years. In this study, however, we have included all relevant WoS document types for a better coverage of the literature of the research topic analyzed. For example, conference meetings and early access papers may well be interesting for the content analysis of the literature under study. Such literature often anticipates important results, which are published later as regular articles. Furthermore, we have included the literature until the date of search for considering the recent rapid growth of the field. Our search retrieved a final publication set of 8,011 papers published until the date of search (July 1, 2021) and dealing with heat waves (#6 of the search query). We have combined this publication set with climate-change-related search terms from a well-proven search query (Haunschild et al. 2016) resulting in 4,588 papers dealing with heat waves in connection with climate change or global warming (# 11 of the search query). Also, we have selected a subset of 2,373 papers dealing with heat waves and mortality (#13 of the search query). The complete WoS search query is given in the appendix.

The final publication set of 8,011 papers dealing with heat waves still contains some non-relevant papers primarily published during the first half of the 20$^{th}$ century, such as some *Nature* papers within the WoS category *Multidisciplinary Sciences*. Since these papers are assigned only to this broad subject category and have no abstracts and no keywords included, they cannot be excluded using the WoS search and refinement functions. We do not expect any bias through these papers, because their keywords do not appear in our maps. Also, they normally contain very few (if any) cited references, which could bias/impact our reference analysis.

**3.2 Networks**

We used the VOSviewer software package (Van Eck and Waltman 2014) to map co-authorship with regard to the countries of authors (88 countries considered) of the papers dealing with heat waves (www.vosviewer.com). The map of the cooperating countries presented is based on the number of joint publications. The distance between two nodes is proportionate to the amount of co-authorship. Hence, largely cooperating countries are positioned closer to each other. The size of the nodes is proportionate to the number of papers published by authors of the specific countries.

The method that we used for revealing the thematic content of the publication set retrieved from the WoS is based on the analysis of keywords. For better standardization, we chose the keywords allocated by the database producer (keywords plus) rather than the author keywords. We also used the VOSviewer for mapping the thematic content of the 104 key papers selected by reference analysis. This map is also based on keywords plus.

The term maps (keywords plus) are based on co-occurrence for positioning the nodes on the maps. The distance between two nodes is proportionate to the co-occurrence of the terms. The size of the nodes is proportionate to the number of papers with a specific keyword. The nodes on the map are assigned by VOSviewer to clusters based on a specific cluster algorithm (the clusters are highlighted in different colors). These clusters identify closely related (frequently co-occurring) nodes, where each node is assigned to only one cluster.

### 3.3 Reference Publication Year Spectroscopy

A bibliometric method called "Reference Publication Year Spectroscopy" (RPYS, Marx et al. 2014) in combination with the tool CRExplorer (http://www.crexplorer.net, Thor et al. 2016) has proven useful for exploring the cited references within a specific publication set, in order to detect the most important publications of the relevant research field (and also the historical roots). In recent years, several studies have been published, in which the RPYS method was basically described and applied (Marx and Bornmann 2014; Marx et al. 2016; Comins and Hussey 2015). In previous studies, Marx et al. have analyzed the roots of research on global warming (Marx et al. 2017a), the emergence of climate change research in combination with tea production (Marx et al. 2017b) and viticulture (Marx et al. 2017c) from a quantitative (bibliometric) perspective. In this study, we determined which references have been most frequently cited by the papers dealing with heat waves.

RPYS is based on the assumption that peers produce a useful database by their publications, in particular by the references cited therein. This database can be analyzed statistically with regard to the works most important for their specific research field. Whereas individual scientists judge their research field more or less subjectively, the overall community can deliver a more objective picture (based on the principle of "the wisdom of the crowds"). The peers effectively ''vote'' via their cited references on which works turned out to be most important for their research field (Bornmann and Marx 2013). RPYS implies a normalization of citation counts (here: reference counts) with regard to the research area and the time of publication, which both impact the probability to be cited frequently. Basically, the citing and cited papers analyzed were published in the same research field and the reference counts are compared with each other only within the same publication year.

RPYS relies on the following observation: the analysis of the publication years of the references cited by all the papers in a specific research topic shows that publication years are not equally represented. Some years occur particularly frequently among the cited references. Such years appear as distinct peaks in the distribution of the reference publication years (i.e., the RPYS spectrogram). The pronounced peaks are frequently based on a few references that are more frequently cited than other references published in the same year. The frequently cited references are – as a rule – of specific significance to the research topic in question (here: heat waves) and the earlier references among them represent its origins and intellectual roots (Marx et al. 2014).

The RPYS changes the perspective of citation analysis from a times cited to a cited reference analysis (Marx and Bornmann 2016). RPYS does not identify the most highly cited papers of the publication set being studied (as is usually done by bibliometric analyses in research evaluation). RPYS aims to mirror the knowledge base of research (here: on heat waves).

With time, the body of scientific literature of many research fields is growing rapidly, particularly in climate change research (Haunschild et al. 2016). The growth rate of highly dynamic research topics such as research related to heat waves is even larger. As a consequence, the number of potentially citable papers is growing substantially. Toward the present, the peaks of individual publications lie over a broad continuum of newer publications and are less numerous and less pronounced. Due to the many publications cited in the more recent years, the proportion of individual highly cited publications in specific reference publication years falls steadily. Therefore, the distinct peaks in an RPYS spectrogram reveal only the most highly cited papers, in particular the earlier references comprising the historical roots. Further inspection and establishing a more entire and representative list of highly cited works requires consulting the reference table provided by the CRExplorer. The most important references within a specific reference publication year can be identified by sorting the cited references according to the reference publication year (RPY) and subsequently according to the number of cited references (N_CR) in a particular publication year.

The selection of important references in RPYS requires the consideration of two opposing trends: (1) the strongly growing number of references per reference publication year and (2) the fall off near present due to the fact that the newest papers had not sufficient time to accumulate higher citation counts. Therefore, we decided to set different limits for the minimum number of cited references for different periods of reference publication years (1950-1999: N_CR≥50, 2000-2014: N_CR≥150, 2015-2020: N_CR≥100). This is somewhat arbitrary, but is necessary in order to adapt and limit the number of cited references to be presented and discussed.

In order to apply RPYS, all cited references (n = 408,247) of 216,932 unique reference variants have been imported from the papers of our publication set on heat waves (n = 8,011). The cited reference publication years range from 1473 to 2021. We removed all references (297 different cited reference variants) with reference publication years prior to 1900. Due to the very low output of heat wave related papers published before 1990, no relevant literature published already in the 19$^{th}$ century can be expected. Also, global warming was no issue before 1900 since the *Little Ice Age* (a medieval cold period) lasted until the 19$^{th}$ century. The references were sorted according to the reference publication year (RPY) and the number of cited references (N_CR) for further inspection.

The CRExplorer offers the possibility to cluster and merge variants of the same cited reference (Thor et al. 2016). We clustered and merged the associated reference variants in our dataset (which are mainly caused by misspelled references) using the corresponding CRExplorer module. Clustering the reference variants via volume and page numbers and subsequently merging aggregated 374 cited references (for more information on using the CRExplorer see “guide and datasets” at www.crexplorer.net).

After clustering and merging, we applied a further cutback: to focus the RPYS on the most-pronounced peaks, we removed all references (n=212,324) with reference counts below 10 (resulting in a final number of 3,937 cited references) for the detection of the most frequently cited works. A minimum reference count of 10 has proved to be reasonable, in particular for early references (Marx et al. 2014; Marx and Bornmann 2014). The cited reference publication years now range from 1932 to 2020.

## 4 Results

In this study, we have considered all WoS document types for a preferably comprehensive coverage of the literature of the research topic analyzed. The vast majority of the papers of our publication set, however, has been assigned to the document types “article” (n=6.738, 84.1 %), “proceedings paper” (n=485, 6.1 %), and “review” (n=395 papers, 4.9 %). Note that some papers belong to more than one document type.

### 4.1 Time-evolution of literature

In Fig. 1, the time-evolution since the year 1990 of the publications dealing with heat waves is shown (there are only 109 pre-1990 publications dealing with heat waves and covered by the WoS).

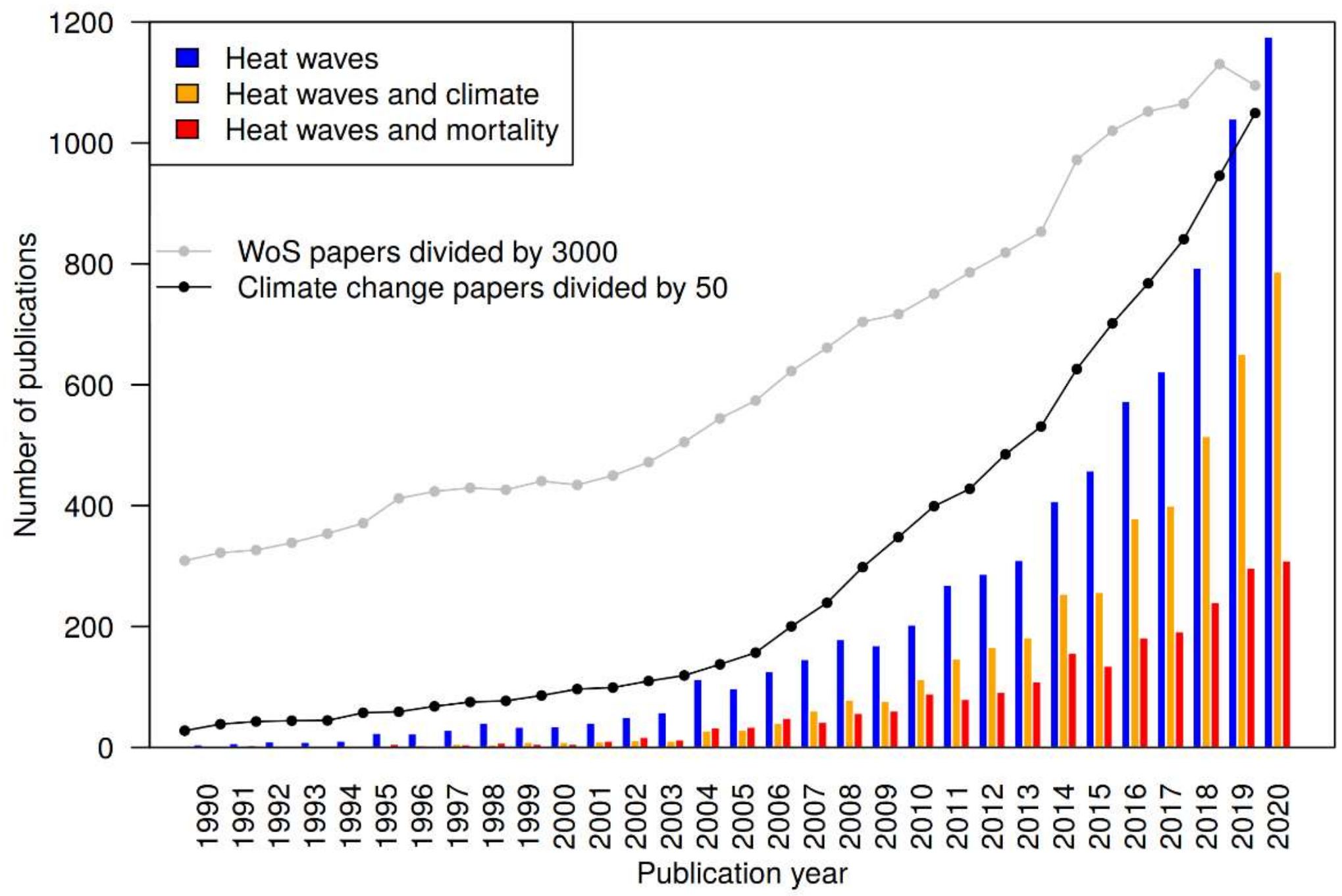

**Fig. 1** Time-evolution of the overall number of heat wave publications, of heat wave publications in connection with climate change, and of heat wave publications in connection with mortality since 1990. For comparison, the overall number of publications (scaled down) in the field of climate change research and the total number of publications covered by the WoS database (scaled down, too) are included.

According to Fig. 1, research dealing with heat waves is a highly dynamic research topic, currently doubling within about 5 years. The number of papers published per year shows a strong increase: since around 2000, the publication output increased by a factor of more than thirty, whereas in the same period the overall number of papers covered by the WoS increased only by a factor of around three. Also, the portion of heat wave papers dealing with climate change increased substantially: from 16.1% in the period 1990-1999 to 25.7% in 2000, reaching 66.9% in 2020. The distinct decrease of the overall number of papers covered by the WoS between 2019 and 2020 might be a result of the Covid-19 pandemic.

With regard to the various impacts of heat waves, excess mortality is one of the most frequently analyzed and discussed issues in the scientific literature (see below). Whereas the subject specific literature on heat waves increased from 2000 to 2020 by a factor of 33.6, literature on heat waves dealing with mortality increased from 2000 to 2020 by a factor of 51.5. The dynamics of the research topic dealing with heat waves is mirrored by the WoS *Citation Report*, which shows the time-evolution of the overall citation impact of the papers of the publication set (not presented). The citation report curve shows no notable citation impact before 2005, corresponding to the increase of the publication rate since about 2003 as shown in Fig. 1.

**4.2 Countries of authors**

In Table 1, the number of papers assigned to the countries of authors with more than 100 publications dealing with heat waves are presented, showing the national part of research activities on this research topic. For comparative purposes, the percentage of overall papers in WoS of each country is shown. As a comparison with the overall WoS, we only considered WoS papers published between 2000 and 2020 because the heat wave literature started to grow substantially around 2000.

**Table 1** Top countries of authors with more than 100 papers dealing with heat waves up to the date of the search.

| Country of authors | #Papers | %Papers heat waves | %Papers overall in WoS |
|---|---|---|---|
| USA | 2081 | 26.0 | 27.4 |
| Australia | 1026 | 12.8 | 3.1 |
| Peoples R China | 965 | 12.0 | 12.1 |
| England | 760 | 9.5 | 6.7 |
| Germany | 737 | 9.2 | 6.3 |
| France | 638 | 8.0 | 4.3 |
| Italy | 536 | 6.7 | 3.9 |
| Spain | 506 | 6.3 | 3.1 |
| Switzerland | 361 | 4.5 | 1.6 |
| Canada | 356 | 4.4 | 4.0 |
| India | 236 | 2.9 | 3.3 |
| Netherlands | 227 | 2.8 | 2.2 |
| South Korea | 206 | 2.6 | 2.5 |
| Sweden | 206 | 2.6 | 1.4 |
| Portugal | 204 | 2.5 | 0.7 |
| Belgium | 176 | 2.2 | 1.2 |
| Japan | 168 | 2.1 | 5.2 |
| Greece | 163 | 2.0 | 0.7 |
| Russia | 149 | 1.9 | 2.1 |
| Poland | 141 | 1.8 | 1.4 |
| Austria | 137 | 1.7 | 0.9 |
| Czech Republic | 130 | 1.6 | 0.7 |
| Denmark | 119 | 1.5 | 0.9 |
| South Africa | 119 | 1.5 | 0.6 |
| Brazil | 116 | 1.4 | 2.1 |
| Scotland | 106 | 1.3 | 1.0 |

The country-specific percentages from Table 1 are visualized in Fig. 2. Selected countries are labeled. Countries with a higher relative percentage of more than two percentage points in heat wave research than in WoS overall output are marked blue (blue circle). Countries with a relative percentage at least twice as high in heat wave research than in overall WoS output are marked green (green cross), whereas countries with a relative percentage at most half as much in heat wave research than in overall WoS output are marked with a yellow cross. Only Japan has a much lower output in heat wave research than in WoS overall output, as indicated by the red circle and yellow cross. Most countries are clustered around the bisecting line and are marked gray (gray circle). China and the USA are outside of the plot region. Both countries are rather close to the bisecting line. Some European countries show a much larger activity in heat wave research than in overall WoS output. Australia shows the largest difference and ratio in output percentages as shown by the blue circle and green cross.

The results mainly follow the expectations of such bibliometric analyses, with one distinct exception: Australia increasingly suffers from extreme heat waves and is comparatively active in heat wave research – compared with its proportion of scientific papers in general. The growth factor of the Australian publication output since 2010 is 8.5, compared to 5.3 for the United States and 3.3 for Germany.

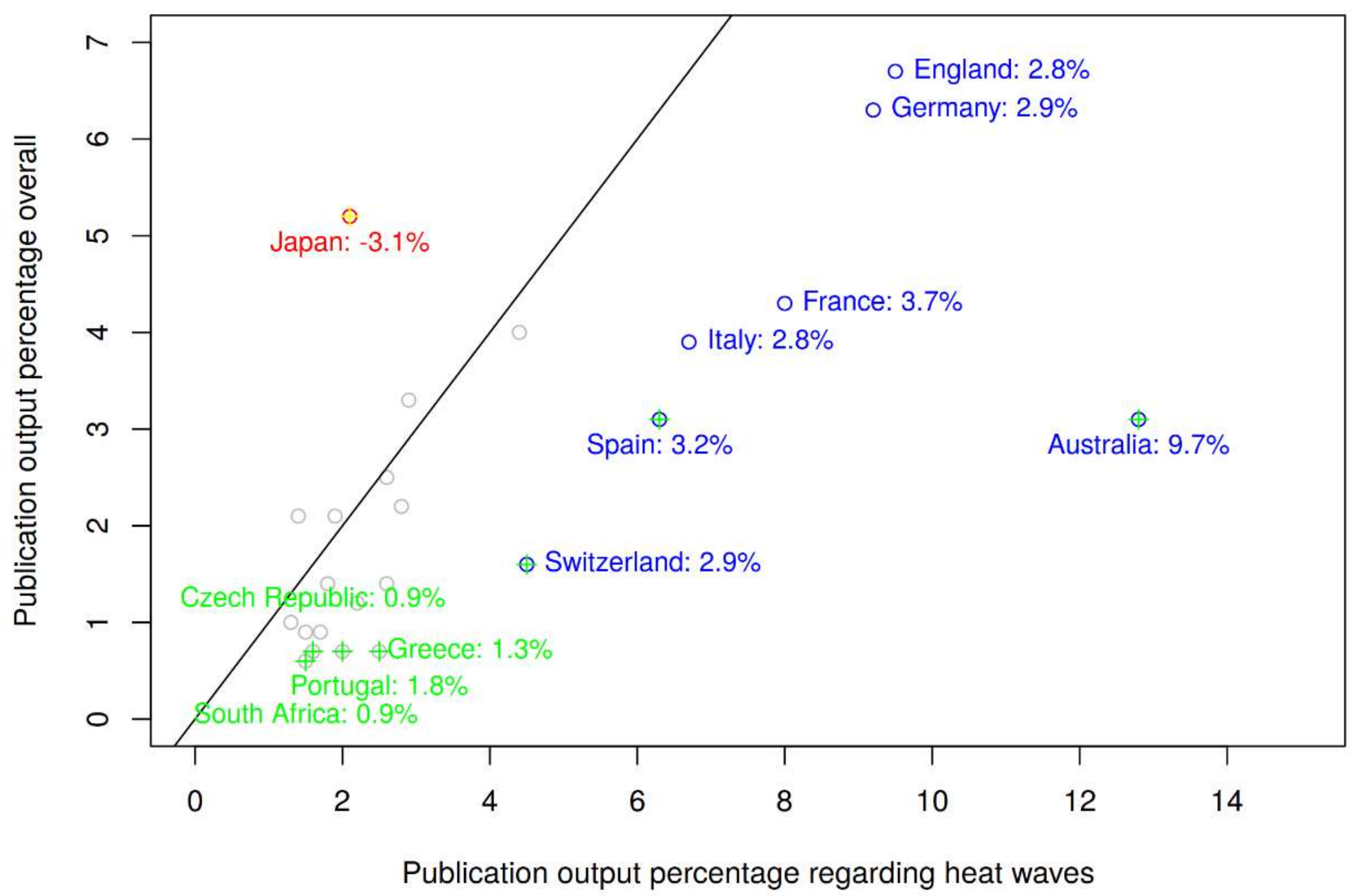

**Fig. 2** Publication percentages of countries in Table 1. Countries with large deviations between heat wave output and overall WoS output are labeled. Countries with an absolute percentage of more than two percentage points higher (lower) in heat wave research than in overall WoS output are marked blue (red). Countries with a relative percentage at least twice as high (at most half as much) in heat wave research than in overall WoS output are marked green (yellow).

Fig. 3 shows the co-authorship network with regard to the countries of authors of the papers dealing with heat waves using the VOSviewer software.

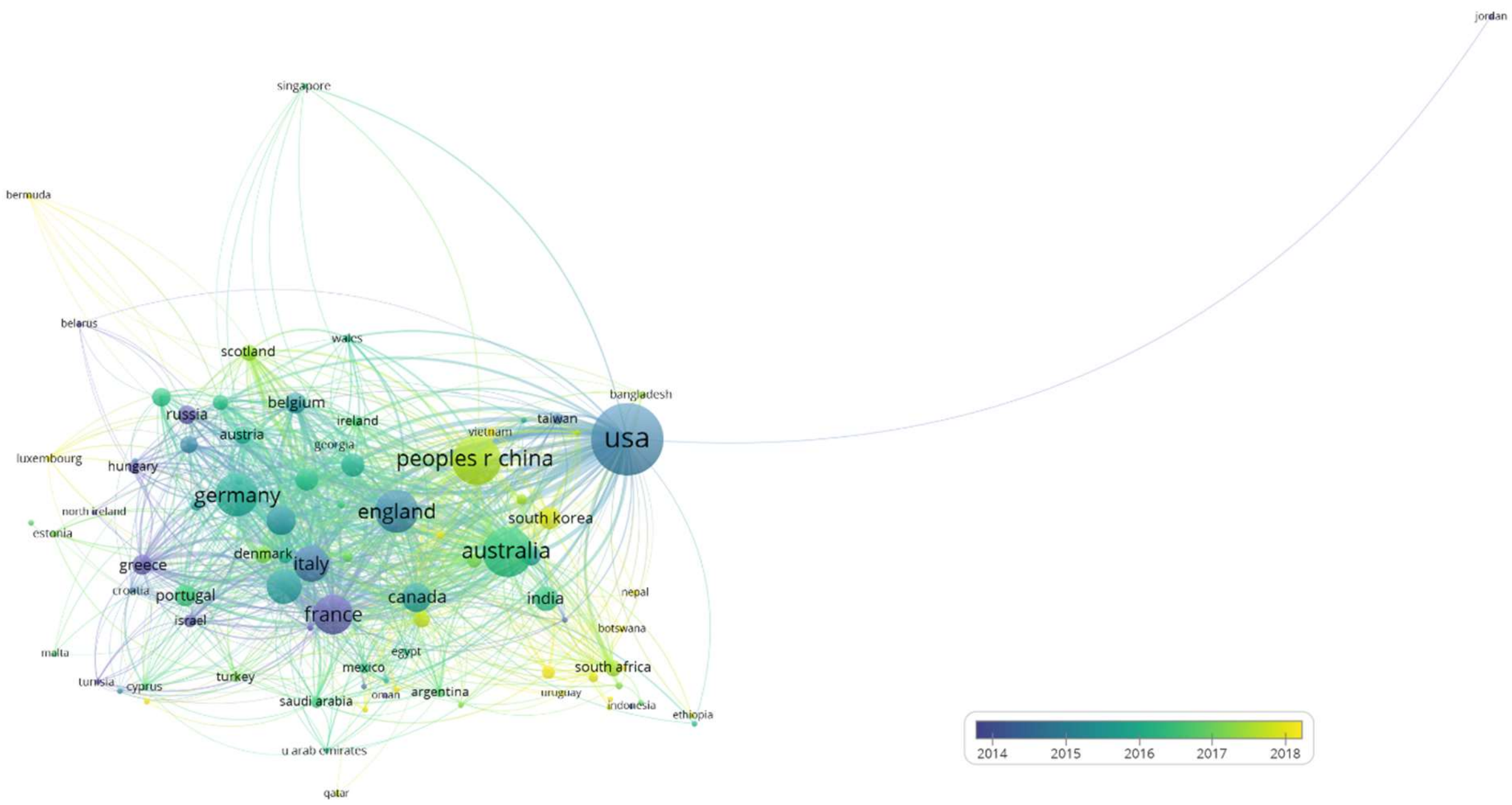

**Fig. 3** Co-authorship overlay map with regard to the countries of authors and their average publication years from the 8,011 papers dealing with heat waves. The minimum number of co-authored publications of a country is 5, papers with more than 25 contributing countries are neglected, of the 135 countries, 89 meet the threshold, 88 out of 89 countries are connected and are considered (one country, Armenia, that is disconnected from the network has been removed). The co-authorship network of a single country can be depicted by clicking on the corresponding node in the interactive map. Readers interested in an in-depth analysis can use VOSviewer interactively and zoom into the clusters via the following URL: https://tinyurl.com/3ywkwv8t

According to Fig. 3 and in accordance with Table 1, the United States are most productive in heat wave research. This is not unexpected, because the US publication output is at the top for most research fields. However, this aside, the United States have been heavily affected by heat wave events and are leading with regard to the emergence of the topic. Australia appears as another major player and is strongly connected with the US publications within the co-authorship network and thus appears as a large node near the US node in the map. Next, the leading European countries England, France, Germany, Italy, and Spain appear.

The overlay version of the map includes the time-evolution of the research activity in the form of coloring of the nodes. The map shows the mean publication year of the publications for each specific author country. As a consequence, the time span of the mean publication years ranges only from 2013 to 2017. Nevertheless, the early activity in France and the United States and the comparatively recent activity in Australia and China, with the European countries in between, becomes clearly visible.

**4.3 Topics of the heat wave literature**

Fig. 4 shows the keywords (keywords plus) map for revealing the thematic content of our publication set using the VOSviewer software. This analysis is based on the complete publication set (n= 8,011). The minimum number of occurrences of keywords is 10; of the 10,964 keywords, 718 keywords met the threshold. For each of the 718 keywords, the total strength of the co-occurrence links with other keywords was calculated. The keywords with the greatest total link strength were selected for presentation in the map.

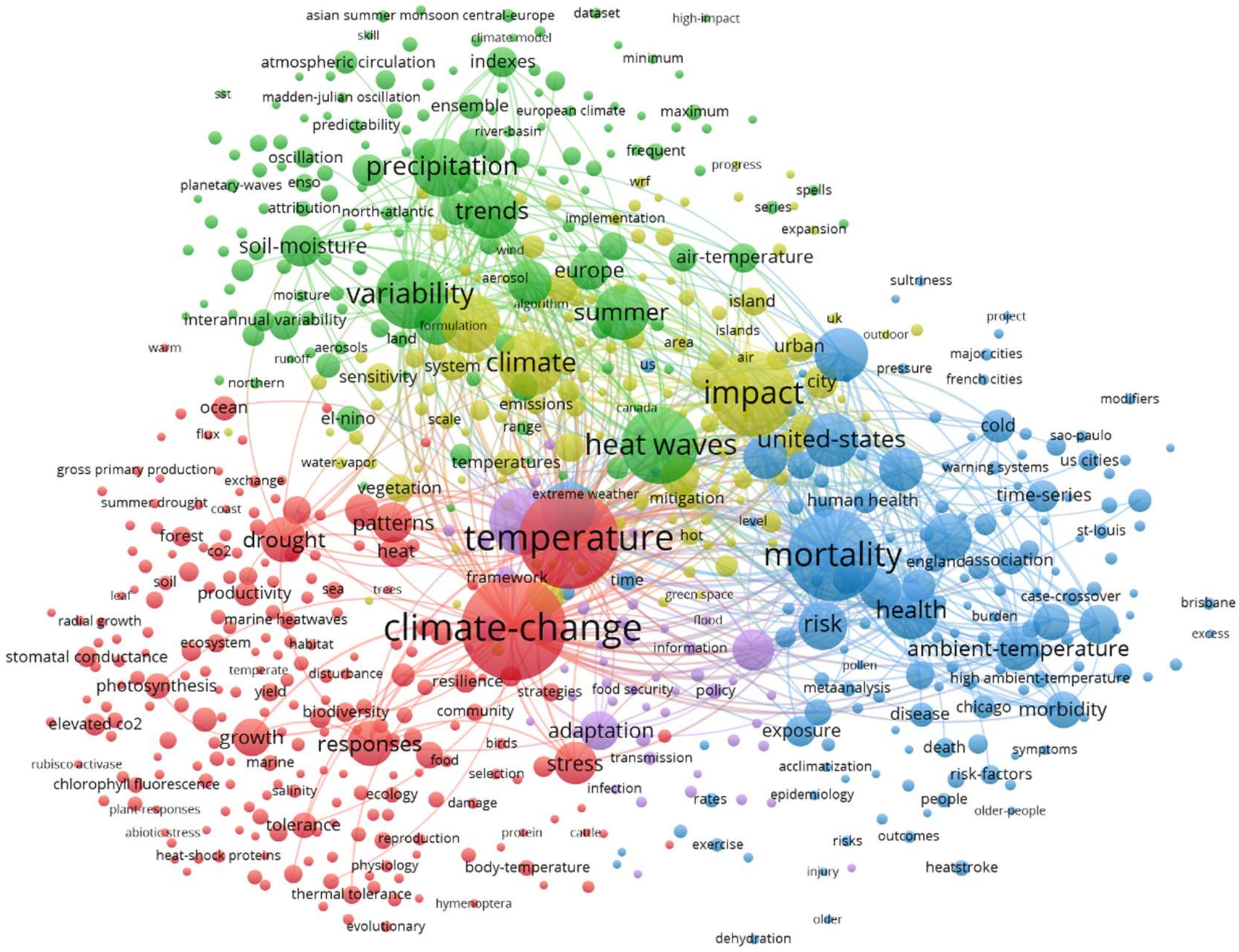

**Fig. 4** Co-occurrence network map of the keywords plus from the 8,011 papers dealing with heat waves for a rough analysis of the thematic content. The minimum number of occurrences of keywords is 10, of the 10,964 keywords, 718 meet the threshold. Readers interested in an in-depth analysis can use VOSviewer interactively and zoom into the clusters via the following URL: https://tinyurl.com/enrdbw

According to Fig. 4, major keywords are: climate change, temperature, mortality, impact, heat waves (searched), and variability. The colored clusters identify closely related (frequently co-occurring) nodes. The keywords marked red roughly originate from fundamental climate change research focused on the hydrological cycle (particularly on drought), the keywords of the green cluster are around heat waves and moisture or precipitation, the keywords marked blue result from research concerning impacts of heat waves on health, the keywords marked yellow are focused on the various other impacts of heat waves, and the keywords of the magenta cluster are around adaptation and vulnerability in connection with heat waves.

The clustering by the VOSviewer algorithm provides basic categorizations, but many related keywords also appear in different clusters. For example, severe heat wave events are marked in different colors. For a better overview of the thematic content of the publications dealing with heat waves, we have assigned the keywords of Fig. 4 (with a minimum number of occurrences of 50) to ten subject categories (each arranged in the order of occurrence):

1. Countries/Regions: United-States, Europe, France, China, Pacific, Australia, London, England
2. Cities: cities, city, US cities, Chicago, communities
3. Events: 2003 heat-wave, 1995 heat-wave
4. Impacts: impact, impacts, air-pollution, drought, soil-moisture, exposure, heat-island, urban, islands, photosynthesis, pollution, heat-island, air-quality, environment, precipitation extremes, biodiversity, emissions
5. Politics: risk, responses, vulnerability, adaptation, management, mitigation, risk-factors, scenarios
6. Biology: vegetation, forest, diversity, stomatal conductance
7. Medicine: mortality, health, stress, deaths, morbidity, hospital admissions, public-health, thermal comfort, population, heat, sensitivity, human health, disease, excess mortality, heat-stress, heat-related mortality, comfort, behavior, death, stroke
8. Climate research: climate change, temperature, climate, model, simulation, energy, projections, simulations, cmip5, ozone, el-nino, parametrization, elevated $CO_2$, models, climate variability, carbon, carbon-dioxide
9. Meteorology: heat waves, variability, precipitation, summer, heat-wave, weather, ambient-temperature, waves, extremes, wave, cold, water, rainfall, circulation, heat, air-temperature, extreme heat, climate extremes, heatwaves, temperature extremes, temperatures, temperature variability, high-temperature, ocean, extreme temperatures, atmospheric circulation, interannual variability, sea-surface temperature, oscillation, surface temperature, surface
10. Broader terms (multi-meaning): trends, events, patterns, growth, performance, time-series, indexes, system, dynamics, association, index, tolerance, productivity, ensemble, resilience, part i, increase, quality, prediction, frequency, particulate matter, future, framework, $20^{th}$-century, time, reanalysis, systems

Although allocated by the database provider, the keywords are not coherent. For example, the same keyword may appear as singular or plural, and complex keywords are written with and without hyphens.

In order to compare the thematic content of the complete publication set with the earlier literature on heat waves, we have analyzed the pre-2000 publications (n=297) separately. Fig. 5 shows the keywords (keywords plus) map for revealing the thematic content of the pre-2000 papers.

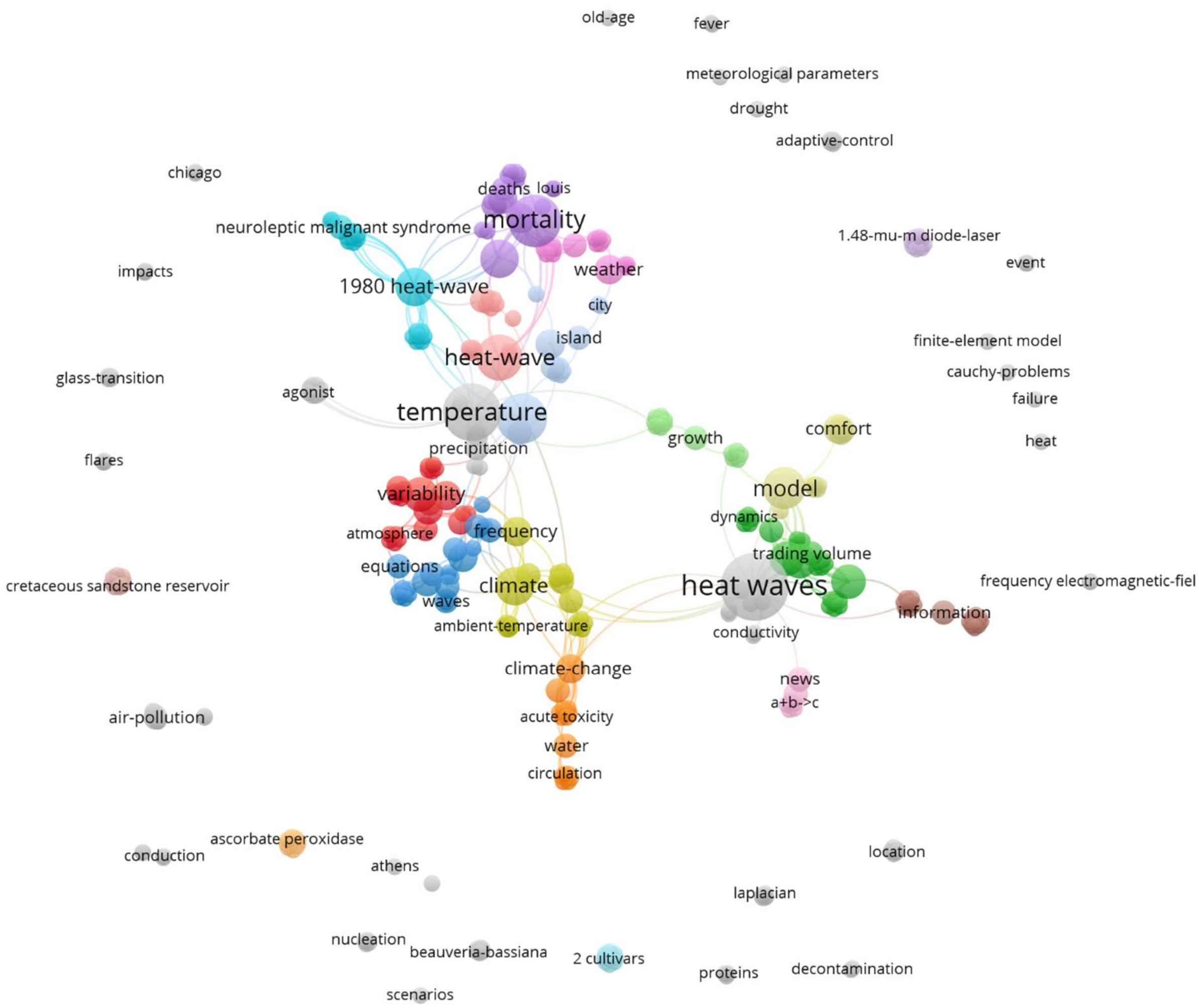

**Fig. 5** Co-occurrence network map of the keywords plus from the 297 pre-2000 papers dealing with heat waves for a rough analysis of the thematic content. The minimum number of occurrences of keywords is 1, of the 389 keywords, 277 keywords are connected, all items are shown. Readers interested in an in-depth analysis can use VOSviewer interactively and zoom into the clusters via the following URL: https://tinyurl.com/u2zzr399

Major nodes in Fig. 5 are: heat waves (searched), temperature, United States, mortality, with climate change appearing only as a smaller node here. Obviously, the connection between heat waves and climate change was not yet pronounced, which can also be seen from Fig. 1. Compared with Fig. 4, the thematic content of the clusters is less clear and the clusters presented in Fig. 5 can hardly be assigned to specific research areas. For a better overview of the thematic content of the early publications dealing with heat waves, we have assigned the connected keywords of Fig. 5 to seven subject categories:

1. Countries/Regions: United-States, Great-Plains
2. Cities: St-Louis, Athens, Chicago
3. Events: 1980 heat-wave, 1995 heat-wave
4. Impacts: impacts, responses, drought, precipitation, comfort, sultriness
5. Climate research: climate, climate change, model, temperature, variability
6. Medicine: cardiovascular deaths, mortality, air pollution
7. Meteorology: atmospheric flow, weather, heat, humidity index

**4.4 Important publications**

Figs. 6-8 show the results of the RPYS analysis performed with the CRExplorer and present the distribution of the number of cited references across the reference publication years. Fig. 6 shows the RPYS spectrogram of the full range of reference publication years since 1925. Fig. 7 presents the spectrogram for the reference publication year period 1950-2000 for better resolving the historical roots. Fig. 8 shows the spectrogram for the period 2000-2020, comprising the cited references from the bulk of the publication set analyzed.

The gray bars (Fig. 6) and red lines (Figs. 7-8) in the graphs visualize the number of cited references per reference publication year. In order to identify those publication years with significantly more cited references than other years, the (absolute) deviation of the number of cited references in each year from the median of the number of cited references in the two previous, the current, and the two following years (t-2; t-1; t; t+1; t+2) is also visualized (blue lines). This deviation from the five-year median provides a curve smoother than the one in terms of absolute numbers. We inspected both curves for the identification of the peak papers.

We use the number of cited references (N_CR) as a measure of the citation impact within the topic specific literature of our publication set. Which papers are most important for the scientific community performing research on heat waves? N_CR should not be confused with the overall number of citations of the papers as given by the WoS citation counts (times cited). These citation counts are based on all citing papers covered by the complete database (rather than a topic-specific publication set) and are usually much higher.

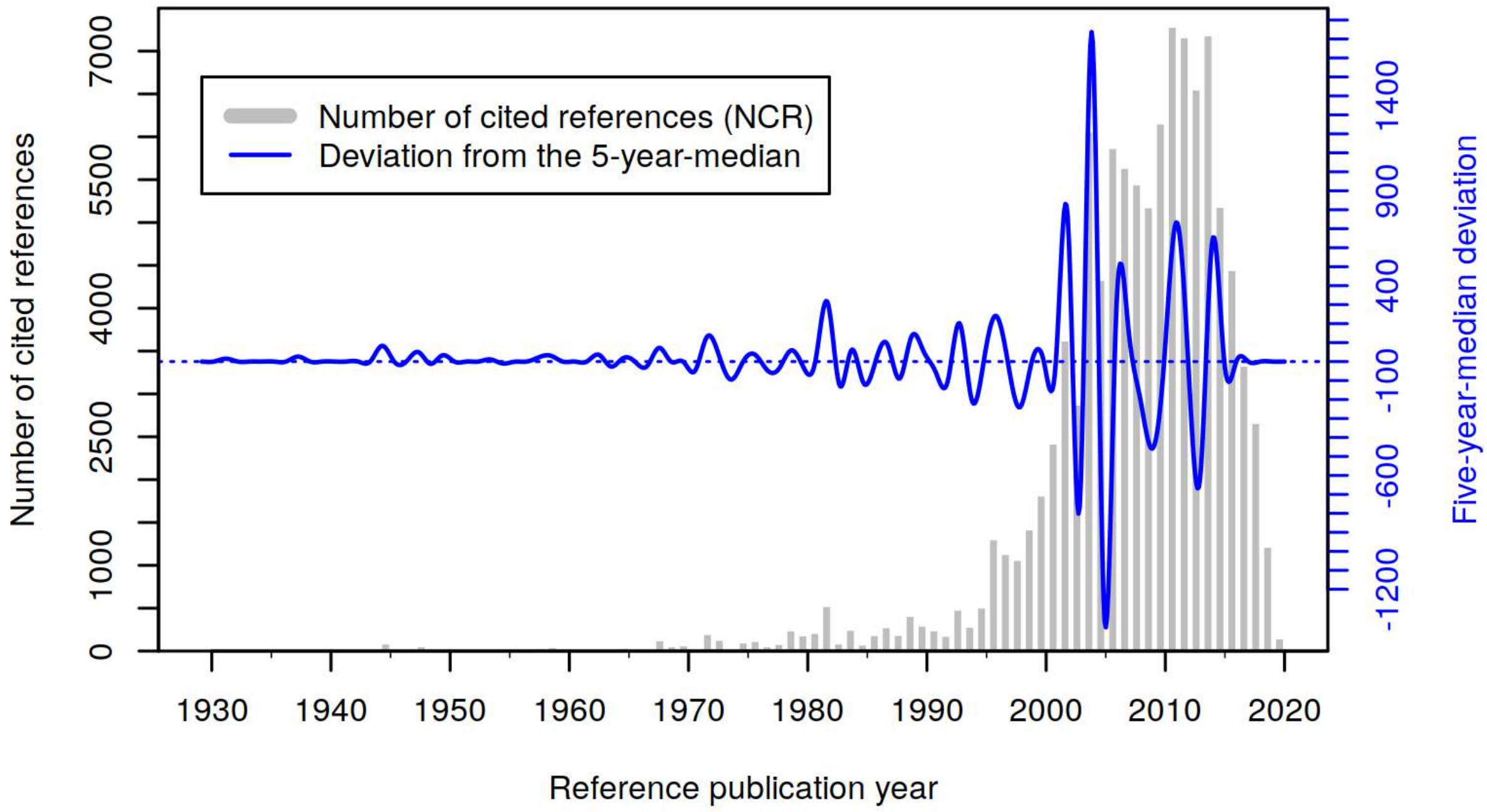

**Fig. 6** Annual distribution of cited references throughout the time period 1925–2020, which have been cited in heat wave related papers (published between 1964 and 2020). Only references with a minimum reference count of 10 are considered.

Applying the selection criteria mentioned above (minimum number of cited references between 50 and 150 in three different periods), 104 references have been selected as key papers (important papers most frequently referenced within the research topic analyzed) and are presented in Table B1 in the appendix. The peak papers corresponding to reference publication years below about 2000 can be seen as the historical roots of the research topic analyzed. Since around 2000, the number of references with the same publication year become increasingly numerous, usually with more than one highly referenced (cited) paper at the top. Although there are comparatively fewer distinct peaks visible in the RPYS spectrogram of Fig. 8, the most frequently referenced papers can easily be identified via the CRE reference listing. Depending on the specific skills and needs (i.e., the expert knowledge and the intended depth of the analysis), the number of top-referenced papers considered as key papers can be defined individually.

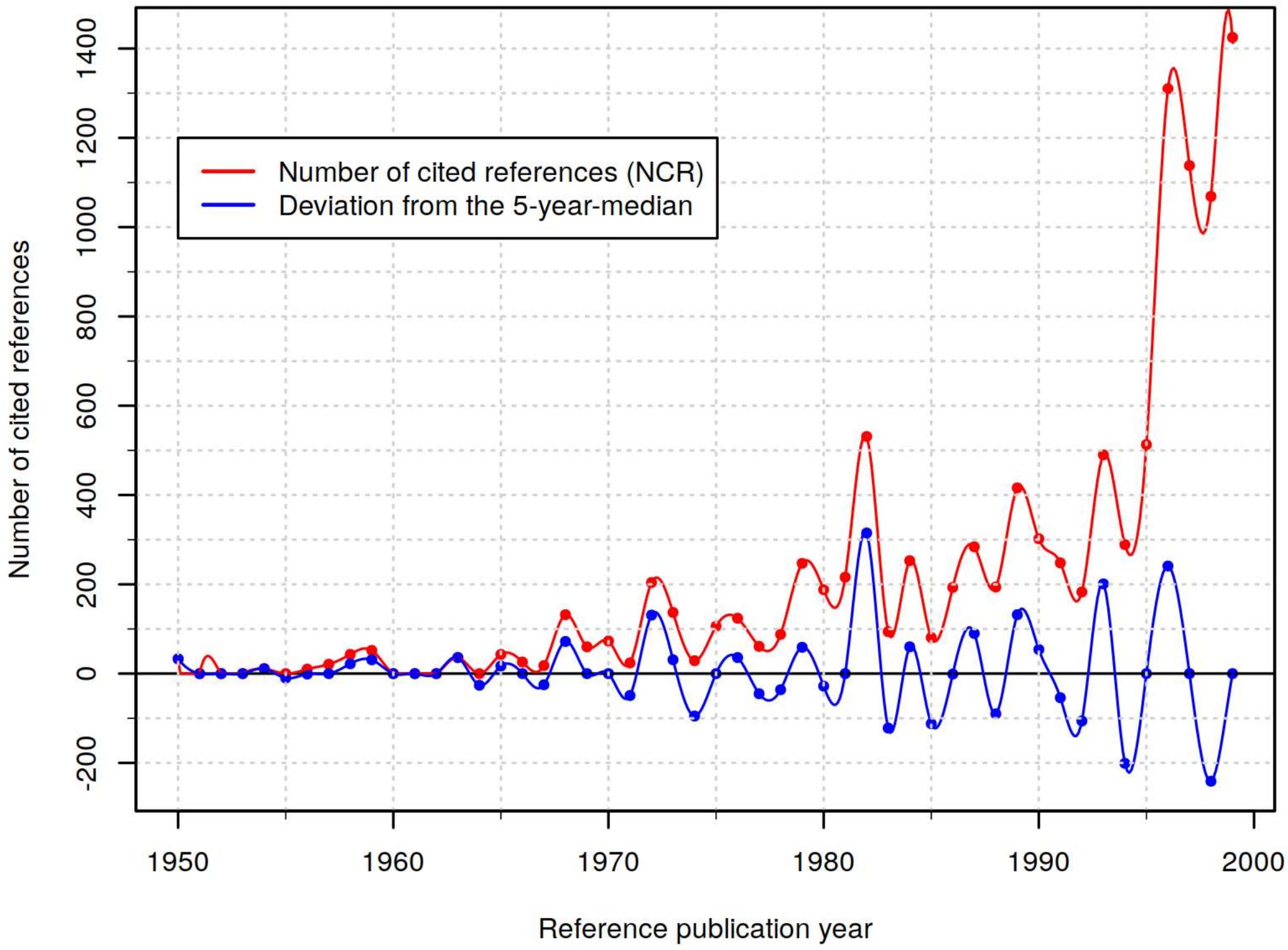

**Fig. 7** Annual distribution of cited references throughout the time period 1950–2000, which have been cited in heat wave related papers (published between 1972 and 2020). Only references with a minimum reference count of 10 are considered.

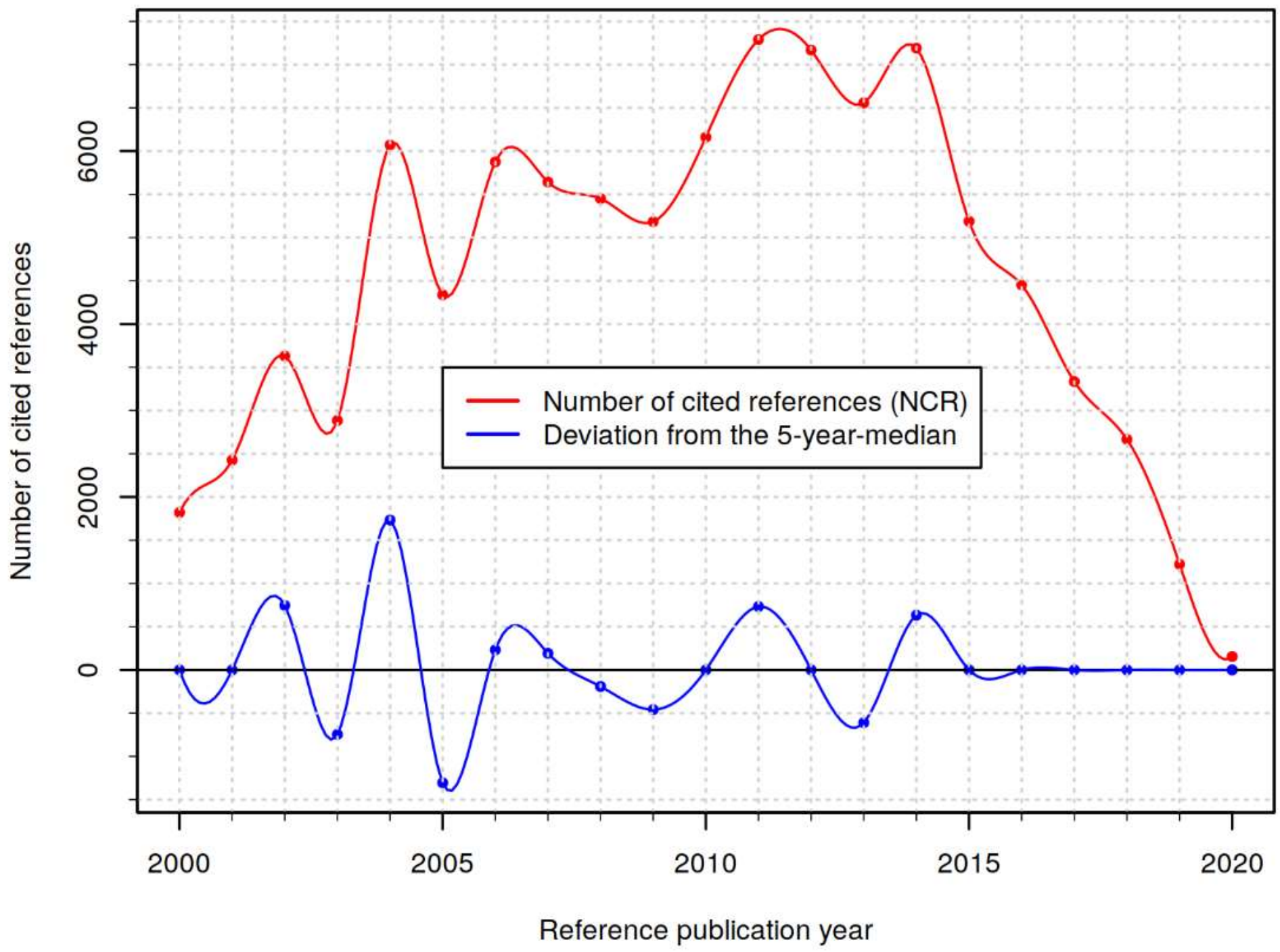

**Fig. 8** Annual distribution of cited references throughout the time period 2000–2020, which have been cited in heat wave related papers (published between 2000 and 2020). Only references with a minimum reference count of 10 are considered.

Table B1 lists the first authors and titles of the 104 key papers selected, their number of cited references (N_CR), and the DOIs for easy access. Some N_CR values are marked by an asterisk, indicating a high value of the N_TOP10 indicator implemented in the CRExplorer. The N_TOP10 indicator value is the number of reference publication years in which a focal cited reference belongs to the 10% most referenced publications. In the case of about half of the cited references in Table B1 (n=58), the N_TOP10 value exceeded a value of 9. The three highest values in our data set are 24, 21, and 20.

Out of the 104 key papers from Table B1, 101 have a DOI of which we found 101 papers in the WoS. Three papers have no DOI but could be retrieved from WoS. The altogether 104 papers were exported and their keywords (keywords plus) were displayed in Fig. 9 for revealing the thematic content of the key papers from the RPYS analysis at a glance.

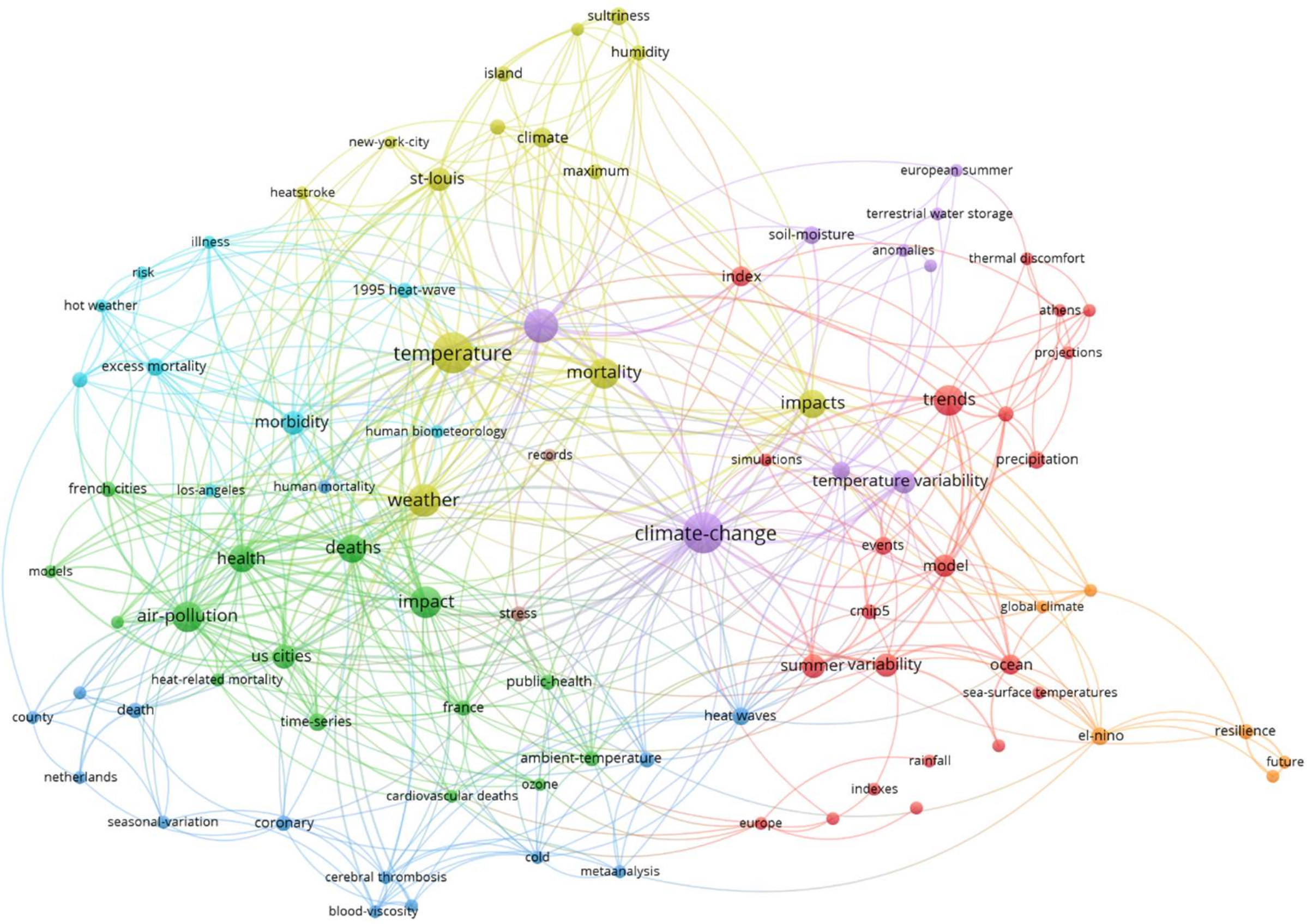

**Fig. 9** Co-occurrence network map of the keywords plus of the 104 key papers dealing with heat waves selected applying RPYS via CRE software and listed in Table B1. The minimum number of occurrences of keywords is 2; of the 310 keywords, 91 meet the threshold. Readers interested in an in-depth analysis can use VOSviewer interactively and zoom into the clusters via the following URL: https://tinyurl.com/4vwpc4t2

Overall, the keywords mapped in Fig. 9 are rather similar to the keywords presented in Fig. 4. Besides climate change, temperature, weather, and air-pollution, the keywords deaths, health, mortality, and United-States appear as the most pronounced terms.

The key papers presented in Table B1 can be categorized as follows: (1) papers dealing with specific heat wave events, (2) the impact of heat waves on human health, (3) heat wave related excess mortality and implications for prevention, (4) the interaction between air pollution and high temperature, (5) circulation pattern and the meteorological basis, (6) future perspectives and risks, (7) climate models, indicators, and statistics.

## 5 Discussion

Today, the hypothesis of a human-induced climate change is no longer abstract but has become a clear fact, at least for the vast majority of the scientific community (IPCC 2014; CSSR 2017; NCA4 2018; and the multitude of references cited therein). The consequences of a warmer climate are already obvious. The rapidly growing knowledge regarding the earth's climate system has revealed the connection between global warming and extreme weather events. Heat waves impact people directly and tangibly and many people are pushing for political actions. Research on heat waves came up with the occurrence of some severe events in the second half of the $20^{th}$ century and was much stimulated by the more numerous, more intense and longer lasting heat waves that have occurred since the beginning of the 21th century.

As already mentioned in the introduction, the more intense and more frequently occurring heat waves cannot be explained solely by natural climate variations but only with human-made climate change. As a consequence, research on heat waves has become embedded into meteorology and climate change research and has aimed to understand the specific connection with global warming. Scientists discuss a weakening of the polar jet stream as a possible reason for an increasing probability for the occurrence of heat waves (e.g., Broennimann et al. 2009; Coumou et al. 2015; Mann 2019). Climate models are used for projections of temperature and rainfall variability in the future, based on various scenarios of greenhouse gas emissions. As a result, the corresponding keywords appear in the maps of Figs. 4 and 9. Also, the application of statistics plays a major role in the papers of our publication set; some of the most-highly referenced (early) papers in Table B1 primarily deal with statistical methods. These methods provide the basis for research on heat waves.

Our analysis shows that research on heat waves has become extremely important in the medical area, since severe heat waves have caused significant excess mortality (e.g., Kalkstein and Davis 1989; Fouillet et al. 2006; Anderson and Bell 2009; Anderson and Bell 2011). Most alarming is that the limit for survivability may be reached at the end of the $21^{st}$ century in many regions of the world due to the fatal combination of rising temperatures and humidity levels (e.g., Pal and Eltahir 2016; Im et al. 2017; Kang and Eltahir 2018). The combination of heat and humidity is measured as the "wet-bulb temperature" (WBT), which is the lowest temperature that can be reached under current ambient conditions by the evaporation of water. At 100% relative humidity, the wet-bulb temperature is equal to the air temperature and is lower at lower humidity levels. For example, an ambient temperature of 46$^{o}$ C and a relative humidity of 50 % corresponds to 35$^{o}$ C WBT, which is the upper limit that can kill even healthy people within hours.

By now, the limit of survivability has almost been reached in some places. However, if global warming is not seriously tackled, deadly heatwaves are anticipated for many regions that have contributed little to climate change. Eastern China and South Asia are particularly at risk, because the annual monsoon brings hot and humid air to these regions. The Gulf Region in the Middle East will also suffer from heatwaves beyond the limit of human survival, indicated by wet-bulb temperatures above 35$^{o}$ C. But to date, only 12 papers have been published on heat waves in connection with wet-bulb temperature (#15 of the search query); no paper was published before 2017. Some papers report excess hospital admissions during heat wave events (e.g., Semenza et al. 1999; Knowlton et al. 2009), with the danger of a temporary capacity overload of local medical systems in the future. Presumably, this will be an increasingly important issue in the future, when more and larger urban areas are affected by heat waves beyond the limit of human survival.

The importance of heat waves for the medical area is underlined by the large portion of papers discussing excess hospital admissions and excess mortality during intense heat wave events, particularly in urban areas with a high population density. As was the case during the boom phase of the Covid-19 pandemic, local medical health care systems may become overstressed by long-lasting heat wave events and thus adaptation strategies are presented and discussed. Finally, the analysis of the keywords in this study reveals the connection of heat wave events with air pollution in urban regions. There seems to be evidence of an interaction between air pollution and high temperatures in the causation of excess mortality (e.g., Katsouyanni et al. 1993). Two more recent papers discuss the global risk of deadly heat (Mora et al. 2017) and the dramatically increasing chance of extremely hot summers since the 2003 European heatwave (Christidis et al. 2015).

Another important topic of the heat wave papers is related to the consequences for agriculture and forestry. Reduced precipitation and soil moisture result in crop failure and put food supplies at risk. Unfortunately, large regions of the world that contribute least to the emission of greenhouse gases are affected most by drought, poor harvests, and hunger. Some more recent papers discuss the increasing probability of marine heatwaves (Oliver et al. 2018) and the consequences for the marine ecosystem (Smale et al. 2019).

The results of this study should be interpreted in terms of its limitations:

(1) We tried to include every relevant heat wave paper in our bibliometric analyses. Our long-standing experience in professional information retrieval has shown, however, that it is sheer impossible to get complete and clean results by search queries against the backdrop of the search functions provided by literature databases like WoS or others. Also, the transition from relevant to non-relevant literature is blurred and is a question of the specific needs. In this study, we used bibliometric methods that are relatively robust with regard to the completeness and precision of the publication sets analyzed. For example, it is an advantage of RPYS that a comparatively small portion of relevant publications (i.e., an incomplete publication set) contains a large amount of the relevant literature as cited references. The number of cited references is indeed lowered as a consequence of an incomplete publication set. However, this does not significantly affect the results, since the reference counts are only used as a relative measure within specific publication years.

Two other limitations of this study also refer to the RPYS of the heat wave paper set:

(2) There are numerous rather highly cited references retrieved by RPYS via CRExplorer but not considered in the listing of Table B1 due to the selection criteria applied. Many of these non-selected papers have N_CR values just below the limits that we have set. Therefore, papers not included in our listing are not per se qualified as much less important or even unimportant.

(3) In the interpretation of cited references counts, one should have in mind that they rely on the ‘popularity’ of a publication being cited in subsequent research. The counts measure impact but not scientific importance or accuracy (Tahamtan and Bornmann 2019). Note that there are many reasons why authors cite publications (Tahamtan and Bornmann 2018), thus introducing a lot of “noise” in the data (this is why RPYS focuses on the cited reference peaks).

Our suggestions for future empirical analysis refer to the impact of the scientific heat wave discourse on social networks and funding of basic research on heat waves around topics driven by political pressure. Whereas this paper focuses on the scientific discourse around heat waves, it would be interesting if future studies were to address the policy relevance of the heat waves research.

**Declarations**

Funding: Not applicable

Conflicts of interest/Competing interests: Not applicable

Availability of data and material: Not applicable

Code availability: Not applicable

Authors' contributions: All authors contributed to the study conception and design. Material preparation, data collection and analysis were performed by Werner Marx, Robin Haunschild, and Lutz Bornmann. The first draft of the manuscript was written by Werner Marx and all authors commented on previous versions of the manuscript. All authors read and approved the final manuscript.

## References

Note: Climate Reports are cited as recommended by CSSR and IPCC.

Anderson BG, Bell ML (2009) Weather-related mortality how heat, cold, and heat waves affect mortality in the United States. Epidemiology 20(2): 205-213. https://doi.org/10.1097/EDE.0b013e318190ee08

Anderson GB, Bell ML (2011) Heat waves in the United States: Mortality risk during heat waves and effect modification by heat wave characteristics in 43 U.S. communities. Environmental Health Perspectives 119(2): 210-218. https://doi.org/10.1289/ehp.1002313

Bornmann L, Marx, W (2013) The wisdom of citing scientists. Journal of the Association for Information Science and Technology 65(6): 1288-1292. https://doi.org/10.1002/asi.23100

Broennimann S, Stickler A, Griesser T, Ewen T, Grant AN, Fischer AM, Schraner M, Peter T, Rozanov E, Ross T (2009) Exceptional atmospheric circulation during the "Dust Bowl". Geophysical Research Letters 36: article number L08802. https://doi.org/10.1029/2009GL037612

Christidis N, Jones G, Stott P (2015) Dramatically increasing chance of extremely hot summers since the 2003 European heatwave. Nature Climate Change 5: 46–50. https://doi.org/10.1038/nclimate2468

Comins JA, Hussey TW (2015) Detecting seminal research contributions to the development and use of the global positioning system by Reference Publication Year Spectroscopy. Scientometrics 104: 575–580. https://doi.org/10.1007/s11192-015-1598-2

Coumou D, Lehmann J, Beckmann J (2015) The weakening summer circulation in the Northern Hemisphere mid-latitudes. Science 348: 324–327. https://doi.org/10.1126/science.1261768

CSSR 2017: Climate Science Special Report: Fourth National Climate Assessment, Volume I [Wuebbles, D.J., D.W. Fahey, K.A. Hibbard, D.J. Dokken, B.C. Stewart, and T.K. Maycock (eds.)]. U.S. Global Change Research Program, Washington, DC, USA, 470 pp., doi: 10.7930/J0J964J6, https://science2017.globalchange.gov/downloads/CSSR2017_FullReport.pdf

Fouillet A, Rey G, Laurent F, Pavillon G, Bellec S, Guihenneuc-Jouyaux C, Clavel J, Jougla E, Hemon D (2006) Excess mortality related to the August 2003 heat wave in France. International Archives of Occupational and Environmental Health 80(1): 16-24. https://doi.org/10.1007/s00420-006-0089-4

Haunschild R, Bornmann L, Marx W (2016) Climate change research in view of bibliometrics. PLoS One 11: e0160393. https://doi.org/10.1371/journal.pone.0160393

Huang QF, Lu YQ (2018) Urban heat island research from 1991 to 2015: a bibliometric analysis. Theoretical and Applied Climatology 131(3-4): 1055-1067. https://doi.org/10.1007/s00704-016-2025-1

Im ES, Pal JS, Eltahir EAB (2017) Deadly heat waves projected in the densely populated agricultural regions of South Asia. Science Advances 3(8): article number e1603322. https://doi.org/10.1126/sciadv.1603322

IPCC 2014: Climate change, Synthesis Report. Contribution of Working Groups I, II and III to the Fifth Assessment Report of the Intergovernmental Panel on Climate Change [Core Writing Team, R.K. Pachauri and L.A. Meyer (eds.)]. IPCC, Geneva, Switzerland, 151 pp. https://www.ipcc.ch/report/ar5/syr/

IPCC 2014: Climate Change, Summary for Policymakers. In: IPCC, 2014: Climate Change 2014: Synthesis Report. Contribution of Working Groups I, II and III to the Fifth Assessment Report of the Intergovernmental Panel on Climate Change [Core Writing Team, R.K. Pachauri and L.A. Meyer (eds.)]. IPCC, Geneva, Switzerland, 151 pp. https://www.ipcc.ch/site/assets/uploads/2018/02/AR5_SYR_FINAL_SPM.pdf

Kalkstein LS, Davis RE (1989) Weather and human mortality – An evaluation of demographic and interregional responses in the United-States. Annals of the Association of American Geographers 79(1): 44-64. https://doi.org/10.1111/j.1467-8306.1989.tb00249.x

Kang S, Eltahir EAB (2018) North China Plain threatened by deadly heatwaves due to climate change and irrigation. Nature Communications 9: article number 2894. https://doi.org/10.1038/s41467-018-05252-y

Katsouyanni K, Pantazopoulou A, Touloumi G, Tselepidaki I, Moustris K, Asimakopoulos D, Poulopoulou G, Trichopoulos D (1993) Evidence for interaction between air-pollution and high-temperature in the causation of excess mortality. Archives of Environmental Health 48(4): 235-242. https://doi.org/10.1080/00039896.1993.9940365

Knowlton K, Rotkin-Ellman M, King G, Margolis HG, Smith D, Solomon G, Trent R, English P (2009) The 2006 California heat wave: Impacts on hospitalizations and emergency department visits. Environmental Health Perspectives 117(1): 61-67. https://doi.org/10.1289/ehp.11594

Mann ME (2019) The weather amplifier: Strange waves in the jet stream foretell a future full of heat waves and floods. Scientific American 320(3): 43-49.

Marx W, Bornmann L (2016) Change of perspective: Bibliometrics from the point of view of cited references. A literature overview on approaches to the evaluation of cited references in bibliometrics. Scientometrics 109(2): 1397–1415. https://doi.org/10.1007/s11192-016-2111-2

Marx W, Haunschild R, Thor A, Bornmann L (2017a) Which early works are cited most frequently in climate change research literature? A bibliometric approach based on reference publication year spectroscopy. Scientometrics 110(1): 335-353. https://doi.org/10.1007/s11192-016-2177-x

Marx W, Haunschild R, Bornmann L (2017b) Climate change and viticulture—A quantitative analysis of a highly dynamic research field. Vitis 56: 35–43. https://doi.org/10.5073/vitis.2017.56.35-43

Marx W, Haunschild R, Bornmann L (2017c) Global warming and tea production – The Bibliometric view on a newly emerging research topic. Climate 5(3): article number 46. https://doi.org/10.3390/cli5030046

Marx W, Bornmann L, Barth A, Leydesdorff L (2014) Detecting the historical roots of research fields by Reference Publication Year Spectroscopy (RPYS). Journal of the Association for Information Science and Technology 65: 751–764. https://doi.org/10.1002/asi.23089

Meehl GA, Tebaldi, C (2004) More intense, more frequent, and longer lasting heat waves in the 21st century. Science 305: 994-997. https://doi.org/10.1126/science.1098704

Mora C, Dousset B, Caldwell I, et al. (2017) Global risk of deadly heat. Nature Climate Change 7: 501–506. https://doi.org/10.1038/nclimate3322

NCA4 2018: Fourth National Climate Assessment, Volume II: Impacts, Risks, and Adaptation in the United States. https://nca2018.globalchange.gov/ NCA 2018 Report-in-Brief: https://nca2018.globalchange.gov/downloads/NCA4_Report-in-Brief.pdf

Oliver ECJ, Donat MG, Burrows MT, et al. (2018) Longer and more frequent marine heatwaves over the past century. Nature Communications 9: 1324. https://doi.org/10.1038/s41467-018-03732-9

Pal JS, Eltahir EAB (2016) Future temperature in southwest Asia projected to exceed a threshold for human adaptability. Nature Climate Change 6(2): 197-200. https://doi.org/10.1038/NCLIMATE2833

Semenza JC, McCullough JE, Flanders WD, McGeehin MA, Lumpkin JR (1999) Excess hospital admissions during the July 1995 heat wave in Chicago. American Journal of Preventive Medicine 16(4): 269-277. https://doi.org/10.1016/S0749-3797(99)00025-2

Smale DA, Wernberg T, Oliver ECJ, et al. Marine heatwaves threaten global biodiversity and the provision of ecosystem services. Nature Climate Change 9: 306–312 (2019). https://doi.org/10.1038/s41558-019-0412-1

Tahamtan I, Bornmann L (2018) Core elements in the process of citing publications: Conceptual overview of the literature. Journal of Informetrics 12: 203-216. https://doi.org/10.1016/j.joi.2018.01.002

Tahamtan I, Bornmann L (2019) What do citation counts measure? An updated review of studies on citations in scientific documents published between 2006 and 2018. Scientometrics 121: 1635–1684. https://doi.org/10.1007/s11192-019-03243-4

Thor A, Marx W, Leydesdorff L, Bornmann L (2016) Introducing CitedReferencesExplorer (CRExplorer): A program for Reference Publication Year Spectroscopy with cited references disambiguation. Journal of Informetrics 10: 503–515. https://doi.org/10.1016/j.joi.2016.02.005

Thor A, Marx W, Leydesdorff L, Bornmann L (2016) New features of CitedReferencesExplorer (CRExplorer). Scientometrics 109: 2049–2051. https://doi.org/10.1007/s11192-016-2082-3

Thor A, Bornmann L, Marx W, Mutz R (2018) Identifying single influential publications in a research field: New analysis opportunities of the CRExplorer. Scientometrics 116(1): 591-608. https://doi.org/10.1007/s11192-018-2733-7

Van Eck NJ, Waltman L (2014) CitNetExplorer: A new software tool for analyzing and visualizing citation networks. Journal of Informetrics 8(4): 802–823. https://doi.org/10.1016/j.joi.2014.07.006

# Appendix A

## WoS search query

| | | |
|---|---|---|
| # 15 | 12 | #14 AND #6<br>*Indexes=SCI-EXPANDED, SSCI, A&HCI, CPCI-S, CPCI-SSH, BKCI-S, BKCI-SSH, ESCI Timespan=All years* |
| # 14 | 1,293 | **TOPIC:** ("wet bulb temperature*" OR WBT)<br>*Indexes=SCI-EXPANDED, SSCI, A&HCI, CPCI-S, CPCI-SSH, BKCI-S, BKCI-SSH, ESCI Timespan=All years* |
| # 13 | 2,373 | #6 AND TS=mortality<br>*Indexes=SCI-EXPANDED, SSCI, A&HCI, CPCI-S, CPCI-SSH, BKCI-S, BKCI-SSH, ESCI Timespan=All years* |
| # 12 | 297 | #2 OR #4<br>**Refined by: DOCUMENT TYPES:** ( ARTICLE OR MEETING ABSTRACT OR CORRECTION OR PROCEEDINGS PAPER OR LETTER OR REVIEW OR NEWS ITEM OR BOOK CHAPTER OR EARLY ACCESS OR EDITORIAL MATERIAL OR BOOK REVIEW ) AND **PUBLICATION YEARS:** ( 1984 OR 1967 OR 1983 OR 1966 OR 1982 OR 1965 OR 1999 OR 1981 OR 1964 OR 1998 OR 1980 OR 1963 OR 1997 OR 1979 OR 1962 OR 1996 OR 1978 OR 1961 OR 1995 OR 1977 OR 1959 OR 1994 OR 1976 OR 1954 OR 1993 OR 1975 OR 1949 OR 1992 OR 1974 OR 1940 OR 1991 OR 1973 OR 1938 OR 1990 OR 1972 OR 1930 OR 1989 OR 1971 OR 1926 OR 1988 OR 1970 OR 1914 OR 1987 OR 1969 OR 1912 OR 1986 OR 1968 OR 1906 OR 1985 )<br>*Indexes=SCI-EXPANDED, SSCI, A&HCI, CPCI-S, CPCI-SSH, BKCI-S, BKCI-SSH, ESCI Timespan=All years* |
| # 11 | 4,588 | #10 AND #6<br>*Indexes=SCI-EXPANDED, SSCI, A&HCI, CPCI-S, CPCI-SSH, BKCI-S, BKCI-SSH, ESCI Timespan=All years* |
| # 10 | 498,361 | #9 OR #8 OR #7<br>*Indexes=SCI-EXPANDED, SSCI, A&HCI, CPCI-S, CPCI-SSH, BKCI-S, BKCI-SSH, ESCI Timespan=All years* |
| # 9 | 191,712 | **TITLE:** (climat* OR palaeoclimat* OR paleoclimat*)<br>*Indexes=SCI-EXPANDED, SSCI, A&HCI, CPCI-S, CPCI-SSH, BKCI-S, BKCI-SSH, ESCI Timespan=All years* |
| # 8 | 121,301 | **TOPIC:** ("global temperature*" OR "global warm*" OR "greenhouse effect" OR "greenhouse gas*" OR "greenhouse warm*")<br>*Indexes=SCI-EXPANDED, SSCI, A&HCI, CPCI-S, CPCI-SSH, BKCI-S, BKCI-SSH, ESCI Timespan=All years* |
| # 7 | 331,878 | **TOPIC:** ("climate chang*" OR "climatic chang*" OR "climate varia*" OR "climatic varia*" OR "climate warm*" OR "climatic warm*")<br>*Indexes=SCI-EXPANDED, SSCI, A&HCI, CPCI-S, CPCI-SSH, BKCI-S, BKCI-SSH, ESCI Timespan=All years* |
| # 6 | 8,011 | #2 OR #4<br>**Refined by: DOCUMENT TYPES:** ( ARTICLE OR MEETING ABSTRACT OR CORRECTION OR PROCEEDINGS PAPER OR LETTER OR REVIEW OR NEWS ITEM OR BOOK CHAPTER OR EARLY ACCESS OR EDITORIAL MATERIAL OR BOOK REVIEW )<br>*Indexes=SCI-EXPANDED, SSCI, A&HCI, CPCI-S, CPCI-SSH, BKCI-S, BKCI-SSH, ESCI Timespan=All years* |
| # 5 | 8,033 | #2 OR #4<br>*Indexes=SCI-EXPANDED, SSCI, A&HCI, CPCI-S, CPCI-SSH, BKCI-S, BKCI-SSH, ESCI Timespan=All years* |
| # 4 | 62 | #3 AND TS=(climat* OR greenhouse OR warming OR atmospher* OR tropospher* OR weather)<br>*Indexes=SCI-EXPANDED, SSCI, A&HCI, CPCI-S, CPCI-SSH, BKCI-S, BKCI-SSH, ESCI Timespan=All years* |
| # 3 | 597 | #1 NOT #2<br>*Indexes=SCI-EXPANDED, SSCI, A&HCI, CPCI-S, CPCI-SSH, BKCI-S, BKCI-SSH, ESCI Timespan=All years* |
| # 2 | 7,971 | **TOPIC:** ("heat wave" OR "heat waves" OR heatwave OR heatwaves OR "hot spell" OR "hot spells")<br>**Refined by:** [excluding] **WEB OF SCIENCE CATEGORIES:** ( NANOSCIENCE NANOTECHNOLOGY OR ASTRONOMY ASTROPHYSICS OR NUCLEAR SCIENCE TECHNOLOGY OR PHYSICS APPLIED OR PHYSICS ATOMIC MOLECULAR CHEMICAL OR PHYSICS CONDENSED MATTER OR PHYSICS FLUIDS PLASMAS OR PHYSICS MATHEMATICAL OR PHYSICS MULTIDISCIPLINARY OR LITERARY REVIEWS OR MECHANICS )<br>*Indexes=SCI-EXPANDED, SSCI, A&HCI, CPCI-S, CPCI-SSH, BKCI-S, BKCI-SSH, ESCI Timespan=All years* |
| # 1 | 8,568 | **TOPIC:** ("heat wave" OR "heat waves" OR heatwave OR heatwaves OR "hot spell" OR "hot spells")<br>*Indexes=SCI-EXPANDED, SSCI, A&HCI, CPCI-S, CPCI-SSH, BKCI-S, BKCI-SSH, ESCI Timespan=All years* |

Date of search: 01.07.2021

## Appendix B

**Table B1** Listing of the key papers (n=104) revealed by RPYS via CRE (RPY = reference publication year, N_CR = number of cited references, Title = title of the cited reference, DOI allows easily to retrieve the full paper via WoS or internet).

| RPY | N_CR | First author | Title | DOI |
|---|---|---|---|---|
| 1945 | 94 | Mann, H.B. | Nonparametric tests against trend | Not available |
| 1968 | 108 | Sen, P.K. | Estimates of regression coefficient based on Kendalls Tau | Not available |
| 1972 | 72 | Schuman, S.H. | Patterns of urban heat-wave deaths and implications for prevention – Data from New York and St-Louis during July, 1966 | 10.1016/0013-9351(72)90020-5 |
| 1973 | 81* | Oke, T.R. | City size and urban heat island | 10.1016/0004-6981(73)90140-6 |
| 1979 | 110* | Steadman, R.G. | The assessment of sultriness. Part I: A temperature-humidity index based on human physiology and clothing science | 10.1175/1520-0450(1979)018<0861:TAOSPI>2.0.CO;2 |
| 1980 | 62 | Berry, J. | Photosynthetic response and adaptation to temperature in higher-plants | 10.1146/annurev.pp.31.060180.002423 |
| 1982 | 175 | Oke, T.R. | The energetic basis of the urban heat island | 10.1002/qj.49710845502 |
| 1982 | 107* | Jones, T.S. | Morbidity and mortality associated with the July 1980 heat-wave in St.-Louis and Kansas-City, MO | 10.1001/jama.247.24.3327 |
| 1982 | 88 | Kilbourne, E.M. | Risk-factors for heat-stroke – A case-control study | 10.1001/jama.247.24.3332 |
| 1984 | 112* | Steadman, R.G. | A universal scale of apparent temperature | 10.1175/1520-0450(1984)023<1674:AUSOAT>2.0.CO;2 |
| 1984 | 52 | Mearns, L.O. | Extreme high-temperature events – Changes in their probabilities with changes in mean temperature | 10.1175/1520-0450(1984)023<1601:EHTECI>2.0.CO;2 |
| 1986 | 59 | Kalkstein, L.S. | An evaluation of summer discomfort in the United-States using a relative climatological index | 10.1175/1520-0477(1986)067<0842:AEOSDI>2.0.CO;2 |
| 1986 | 53 | Keatinge, W.R. | Increased platelet and red-cell counts, blood-viscosity, and plasma-cholesterol levels during heat stress, and mortality from coronary and cerebral thrombosis | 10.1016/0002-9343(86)90348-7 |
| 1987 | 71 | Mayer, H. | Thermal comfort of man in different urban environments | 10.1007/BF00866252 |
| 1989 | 87* | Kalkstein, L.S. | Weather and human mortality – An evaluation of demographic and interregional responses in the United-States | 10.1111/j.1467-8306.1989.tb00249.x |
| 1989 | 65 | Dudhia, J. | Numerical study of convection observed during the winter monsoon experiment using a mesoscale two-dimensional model | 10.1175/1520-0469(1989)046<3077:NSOCOD>2.0.CO;2 |
| 1989 | 53 | Joseph, D.D. | Heat waves | 10.1103/RevModPhys.61.41 |
| 1992 | 85* | Katz, R.W. | Extreme events in a changing climate – Variability is more important than averages | 10.1007/BF00139728 |
| 1993 | 74* | Kunst, A.E. | Outdoor air temperature and mortality in the Netherlands: A time-series analysis | 10.1093/oxfordjournals.aje.a116680 |
| 1993 | 52 | Katsouyanni, K. | Evidence for interaction between air-pollution and high-temperature in the causation of excess mortality | 10.1080/00039896.1993.9940365 |
| 1995 | 52 | Sartor, F. | Temperature, ambient ozone levels, and mortality during summer, 1994, in Belgium | 10.1006/enrs.1995.1054 |
| 1996 | 368* | Semenza, J.C. | Heat-related deaths during the July 1995 heat wave in Chicago | 10.1056/NEJM199607113350203 |
| 1996 | 327* | Kalnay, E. | The NCEP/NCAR 40-year reanalysis project | 10.1175/1520-0477(1996)077<0437:TNYRP>2.0.CO;2 |
| 1996 | 106* | Changnon, S.A. | Impacts and responses to the 1995 heat wave: A call to action | 10.1175/1520-0477(1996)077<1497:IARTTH>2.0.CO;2 |
| 1996 | 84 | Kunkel, K.E. | The July 1995 heat wave in the midwest: A climatic perspective and critical weather factors | 10.1175/1520-0477(1996)077<1507:TJHWIT>2.0.CO;2 |
| 1996 | 63 | Kalkstein, L.S. | The Philadelphia hot weather-health watch warning system: Development and application, summer 1995 | 10.1175/1520-0477(1996)077<1519:TPHWHW>2.0.CO;2 |
| 1997 | 180* | Whitman, S. | Mortality in Chicago attributed to the July 1995 heat wave | 10.2105/AJPH.87.9.1515 |
| 1997 | 116* | Kalkstein, L.S. | An evaluation of climate/mortality relationships in large US cities and the possible impacts of a climate change | 10.1289/ehp.9710584 |
| 1997 | 105* | Karl, T.R. | The 1995 Chicago heat wave: How likely is a recurrence? | 10.1175/1520-0477(1997)078<1107:TCHWHL>2.0.CO;2 |
| 1997 | 74 | Mlawer, E.J. | Radiative transfer for inhomogeneous atmospheres: RRTM, a validated correlated-k model for the longwave | 10.1029/97JD00237 |
| 1997 | 72 | Keatinge, W.R. | Cold exposure and winter mortality from ischaemic heart disease, cerebrovascular disease, respiratory disease, and all causes in warm and cold regions of Europe | Not available |
| 1997 | 53 | Easterling, D.R. | Maximum and minimum temperature trends for the globe | 10.1126/science.277.5324.364 |
| 1997 | 53 | Mantua, N.J. | A Pacific interdecadal climate oscillation with impacts on salmon production | 10.1175/1520-0477(1997)078<1069:APICOW>2.0.CO;2 |
| 1997 | 50 | Ballester, F. | Mortality as a function of temperature. A study in Valencia, Spain, 1991-1993 | 10.1093/ije/26.3.551 |
| 1998 | 105* | Rooney, C. | Excess mortality in England and Wales, and in Greater London, during the 1995 heatwave | 10.1136/jech.52.8.482 |
| 1998 | 81* | Dematte, J.E. | Near-fatal heat stroke during the 1995 heat wave in Chicago | 10.7326/0003-4819-129-3-199808010-00001 |
| 1998 | 55 | Smoyer, K.E. | A comparative analysis of heat waves and associated mortality in St. Louis, Missouri - 1980 and 1995 | 10.1007/s004840050082 |
| 1998 | 53 | Gaffen, D.J. | Increased summertime heat stress in the US | 10.1038/25030 |
| 1999 | 200* | Semenza, J.C. | Excess hospital admissions during the July 1995 heat wave in Chicago | 10.1016/S0749-3797(99)00025-2 |
| 1999 | 91 | Hoppe, P. | The physiological equivalent temperature - A universal index for the biometeorological assessment of the thermal environment | 10.1007/s004840050118 |
| 1999 | 56 | Matzarakis, A. | Applications of a universal thermal index: physiological equivalent temperature | 10.1007/s004840050119 |
| 1999 | 52* | Kunkel, K.E. | Temporal fluctuations in weather and climate extremes that cause economic and human health impacts: A review | 10.1175/1520-0477(1999)080<1077:TFIWAC>2.0.CO;2 |
| 2000 | 275* | Easterling, D.R. | Climate extremes: observations, modelling, and impacts | 10.1126/science.289.5487.2068 |
| 2001 | 277* | Robinson, P.J. | On the definition of a heat wave | 10.1175/1520-0450(2001)040<0762:OTDOAH>2.0.CO;2 |
| 2001 | 203* | Hynen, M.M.T.E. | The impact of heat waves and cold spells on mortality rates in the Dutch population | 10.2307/3454704 |
| 2002 | 286* | Basu, R. | Relation between elevated ambient temperature and mortality: A review of the epidemiologic evidence | 10.1093/epirev/mxf007 |
| 2002 | 279* | Curriero, F.C. | Temperature and mortality in 11 cities of the eastern United States | 10.1093/aje/155.1.80 |
| 2002 | 233* | Frich, P. | Observed coherent changes in climatic extremes during the second half of the twentieth century | 10.3354/cr019193 |
| 2002 | 156* | Bouchama, A. | Medical progress - Heat stroke | 10.1056/NEJMra011089 |
| 2004 | 1099* | Meehl, G.A. | More intense, more frequent, and longer lasting heat waves in the $21^{st}$ century | 10.1126/science.1098704 |
| 2004 | 562* | Schaer, C. | The role of increasing temperature variability in European summer heatwaves | 10.1038/nature02300 |
| 2004 | 326* | Stott, P.A. | Human contribution to the European heatwave of 2003 | 10.1038/nature03089 |
| 2004 | 232* | Beniston, P.M. | The 2003 heat wave in Europe: A shape of things to come? An analysis based on Swiss climatological data and model simulations | 10.1029/2003GL018857 |
| 2004 | 209* | Black, E. | Factors contributing to the summer 2003 European heatwave | 10.1256/wea.74.04 |
| 2004 | 180* | Luterbacher, J. | European seasonal and annual temperature variability, trends, and extremes since 1500 | 10.1126/science.1093877 |
| 2004 | 152 | Kovats, R.S. | Contrasting patterns of mortality and hospital admissions during hot weather and heat waves in Greater London, UK | 10.1136/oem.2003.012047 |
| 2005 | 313* | Ciais, P. | Europe-wide reduction in primary productivity caused by the heat and drought in 2003 | 10.1038/nature03972 |
| 2005 | 161* | Patz, J.A. | Impact of regional climate change on human health | 10.1038/nature04188 |
| 2005 | 157* | Conti, S. | Epidemiologic study of mortality during the Summer 2003 heat wave in Italy | 10.1016/j.envres.2004.10.009 |
| 2006 | 298* | Fouillet, A. | Excess mortality related to the August 2003 heat wave in France | 10.1007/s00420-006-0089-4 |
| 2006 | 257* | Alexander, L.V. | Global observed changes in daily climate extremes of temperature and precipitation | 10.1029/2005JD006290 |
| 2006 | 190* | Seneviratne, S.I. | Land-atmosphere coupling and climate change in Europe | 10.1038/nature05095 |
| 2006 | 180* | Vandentorren, S. | August 2003 heat wave in France: Risk factors for death of elderly people living at home | 10.1093/eurpub/ckl063 |
| 2006 | 177* | Hajat, S. | Impact of high temperatures on mortality - Is there an added heat wave effect? | 10.1097/01.ede.0000239688.70829.63 |
| 2006 | 153* | Kovats, R.S. | Heatwaves and public health in Europe | 10.1093/eurpub/ckl049 |
| 2007 | 238* | Fischer, E.M. | Soil moisture - Atmosphere interactions during the 2003 European summer heat wave | 10.1175/JCLI4288.1 |
| 2007 | 156* | Della-Marta, P.M. | Doubled length of western European summer heat waves since 1880 | 10.1029/2007JD008510 |
| 2008 | 372* | Robine, J.M. | Death toll exceeded 70,000 in Europe during the summer of 2003 | 10.1016/j.crvi.2007.12.001 |
| 2008 | 287* | Kovats, R.S. | Heat stress and public health: A critical review | 10.1146/annurev.publhealth.29.020907.090843 |
| 2008 | 178* | Baccini, M. | Heat effects on mortality in 15 European cities | 10.1097/EDE.0b013e318176bfcd |
| 2008 | 175* | Luber, G. | Climate change and extreme heat events | 10.1016/j.amepre.2008.08.021 |
| 2008 | 160* | Fouillet, A. | Has the impact of heat waves on mortality changed in France since the European heat wave of summer 2003? A study of the 2006 heat wave | 10.1093/ije/dym253 |
| 2009 | 339* | Anderson, Brooke G. | Weather-related mortality: How heat, cold, and heat waves affect mortality in the United States | 10.1097/EDE.0b013e318190ee08 |
| 2009 | 212* | Basu, R. | High ambient temperature and mortality: A review of epidemiologic studies from 2001 to 2008 | 10.1186/1476-069X-8-40 |
| 2009 | 205* | Knowlton, K. | The 2006 California heat wave: Impacts on hospitalizations and emergency department visits | 10.1289/ehp.11594 |
| 2010 | 282* | Fischer, E.M. | Consistent geographical patterns of changes in high-impact European heatwaves | 10.1038/NGEO866 |
| 2010 | 208* | D'Ippoliti, D. | The impact of heat waves on mortality in 9 European cities: results from the EuroHEAT project | 10.1186/1476-069X-9-37 |
| 2010 | 208* | Seneviratne, S.I. | Investigating soil moisture-climate interactions in a changing climate: A review | 10.1016/j.earscirev.2010.02.004 |
| 2010 | 196* | Garcia-Herrera, R. | A review of the European summer heat wave of 2003 | 10.1080/10643380802238137 |
| 2011 | 381* | Dee, D.P. | The ERA-Interim reanalysis: configuration and performance of the data assimilation system | 10.1002/qj.828 |
| 2011 | 355* | Barriopedro, D. | The hot summer of 2010: Redrawing the temperature record map of Europe | 10.1126/science.1201224 |
| 2011 | 303* | Anderson, G.B. | Heat waves in the United States: Mortality risk during heat waves and effect modification by heat wave characteristics in 43 U.S. communities | 10.1289/ehp.1002313 |
| 2011 | 198* | Dole, R. | Was there a basis for anticipating the 2010 Russian heat wave? | 10.1029/2010GL046582 |
| 2011 | 154* | Gasparrini, A. | The Impact of Heat Waves on Mortality | 10.1097/EDE.0b013e3181fdcd99 |
| 2012 | 243 | Taylor, K.E. | An overview of CMIP5 and the experiment design | 10.1175/BAMS-D-11-00094.1 |
| 2012 | 236 | Perkins, S.E. | Increasing frequency, intensity and duration of observed global heatwaves and warm spells | 10.1029/2012GL053361 |
| 2012 | 204 | Coumou, D. | A decade of weather extremes | 10.1038/NCLIMATE1452 |
| 2013 | 289 | Perkins, S.E. | On the measurement of heat waves | 10.1175/JCLI-D-12-00383.1 |
| 2013 | 159 | Wernberg, T. | An extreme climatic event alters marine ecosystem structure in a global biodiversity hotspot | 10.1038/NCLIMATE1627 |
| 2014 | 197 | Miralles, D.G. | Mega-heatwave temperatures due to combined soil desiccation and atmospheric heat accumulation | 10.1038/NGEO2141 |
| 2014 | 171 | Russo, S. | Magnitude of extreme heat waves in present climate and their projection in a warming world | 10.1002/2014JD022098 |
| 2015 | 182 | Gasparrini, A. | Mortality risk attributable to high and low ambient temperature: A multicountry observational study | 10.1016/S0140-6736(14)62114-0 |
| 2015 | 168 | Perkins, S.E. | A review on the scientific understanding of heatwaves - Their measurement, driving mechanisms, and changes at the global scale | 10.1016/j.atmosres.2015.05.014 |
| 2015 | 159 | Russo, S. | Top ten European heatwaves since 1950 and their occurrence in the coming decades | 10.1088/1748-9326/10/12/124003 |
| 2015 | 131 | Bond, N.A. | Causes and impacts of the 2014 warm anomaly in the NE Pacific | 10.1002/2015GL063306 |
| 2015 | 111 | Christidis, N. | Dramatically increasing chance of extremely hot summers since the 2003 European heatwave | 10.1038/NCLIMATE2468 |
| 2016 | 197 | Hobday, A.J. | A hierarchical approach to defining marine heatwaves | 10.1016/j.pocean.2015.12.014 |
| 2016 | 121 | Wernberg, T. | Climate-driven regime shift of a temperate marine ecosystem | 10.1126/science.aad8745 |
| 2016 | 113 | Di Lorenzo, E.. | Multi-year persistence of the 2014/15 North Pacific marine heatwave | 10.1038/NCLIMATE3082 |
| 2017 | 125 | Mora, C. | Global risk of deadly heat | 10.1038/NCLIMATE3322 |
| 2017 | 112 | Hughes, T.P. | Global warming and recurrent mass bleaching of corals | 10.1038/nature21707 |
| 2018 | 209 | Oliver, E.C.J. | Longer and more frequent marine heatwaves over the past century | 10.1038/s41467-018-03732-9 |
| 2018 | 124 | Frolicher, T.L. | Marine heatwaves under global warming | 10.1038/s41586-018-0383-9 |
| 2019 | 126 | Smale, D.A. | Marine heatwaves threaten global biodiversity and the provision of ecosystem services | 10.1038/s41558-019-0412-1 |

*: N_TOP10 >9, the N_TOP10 indicator is the number of reference publication years in which a focal cited reference belongs to the 10% most referenced publications.